\documentclass[sigconf, nonacm]{acmart}
\usepackage[linesnumbered,vlined,ruled]{algorithm2e}
\usepackage{xcolor}
\usepackage{mdframed}
\usepackage{url}
\usepackage{graphicx}
\usepackage{pifont}
\usepackage{appendix}
\usepackage{multirow}
\usepackage{rotating}
\usepackage{booktabs}
\usepackage{mathtools}
\usepackage{caption}
\usepackage{subcaption}
\usepackage{amsmath}
\usepackage{balance}
\DeclarePairedDelimiter{\ceil}{\lceil}{\rceil}

\newtheorem{proposition}{Proposition}

\newcommand\vldbdoi{XX.XX/XXX.XX}

\newcommand\vldbpagestyle{plain} 

\begin{document}
\title{FlowWalker: A Memory-efficient and High-performance GPU-based Dynamic Graph Random Walk Framework}

%%
%% The "author" command and its associated commands are used to define the authors and their affiliations.

\author{Junyi Mei$^1$, Shixuan Sun$^1$, Chao Li$^1$, Cheng Xu$^1$, Cheng Chen$^2$, Yibo Liu$^1$, Jing Wang$^1$, Cheng Zhao$^2$, Xiaofeng Hou$^1$, Minyi Guo$^1$, Bingsheng He$^3$, Xiaoliang Cong$^2$}

\affiliation{%
  \institution{$^1$Shanghai Jiao Tong University, $^2$ByteDance Inc., $^3$National University of Singapore}
}

\email{meijunyi@sjtu.edu.cn, sunshixuan@sjtu.edu.cn, lichao@cs.sjtu.edu.cn, jerryxu@sjtu.edu.cn}
\email{chencheng.sg@bytedance.com,liuyib@sjtu.edu.cn, jing618@sjtu.edu.cn, zhaocheng.127@bytedance.com}
\email{hou-xf@cs.sjtu.edu.cn, guo-my@cs.sjtu.edu.cn, hebs@comp.nus.edu.sg, congxiaoliang@bytedance.com}

% \begin{abstract}
\begin{abstract}
  Dynamic graph random walk (DGRW) emerges as a practical tool for capturing structural relations within a graph.
  Effectively executing DGRW on GPU presents certain challenges. First, existing sampling methods demand a pre-processing buffer, causing substantial space complexity. Moreover, the power-law distribution of graph vertex degrees introduces workload imbalance issues, rendering DGRW embarrassed to parallelize.
  In this paper, we propose Flow\-Walker, a GPU-based dynamic graph random walk framework. Flow\-Walker implements an efficient parallel sampling method to fully exploit the GPU parallelism and reduce space complexity. Moreover, it employs a sampler-centric paradigm alongside a dynamic scheduling strategy to handle the huge amounts of walking queries. Flow\-Walker stands as a memory-efficient framework that requires no auxiliary data structures in GPU global memory.
  We examine the performance of Flow\-Walker extensively on ten datasets, and experiment results show that Flow\-Walker achieves up to $752.2\times$, $72.1\times$, and $16.4\times$ speedup compared with existing CPU, GPU, and FPGA random walk frameworks, respectively.
  Case study shows that Flow\-Walker diminishes random walk time from 35\% to 3\% in a pipeline of Byte\-Dance friend recommendation GNN training. The source code of FlowWalker can be found at \url{https://github.com/junyimei/flowwalker-artifact}.
\end{abstract}

\maketitle


\pagestyle{\vldbpagestyle}

% \begingroup\small\noindent\raggedright\textbf{PVLDB Reference Format:}\\
% \vldbauthors. \vldbtitle. PVLDB, \vldbvolume(\vldbissue): \vldbpages, \vldbyear.\\
% \href{https://doi.org/\vldbdoi}{doi:\vldbdoi}
% \endgroup
% \begingroup
% \renewcommand\thefootnote{}\footnote{\noindent
%   This work is licensed under the Creative Commons BY-NC-ND 4.0 International License. Visit \url{https://creativecommons.org/licenses/by-nc-nd/4.0/} to view a copy of this license. For any use beyond those covered by this license, obtain permission by emailing \href{mailto:info@vldb.org}{info@vldb.org}. Copyright is held by the owner/author(s). Publication rights licensed to the VLDB Endowment. \\
%   \raggedright Proceedings of the VLDB Endowment, Vol. \vldbvolume, No. \vldbissue\ %
%   ISSN 2150-8097. \\
%   \href{https://doi.org/\vldbdoi}{doi:\vldbdoi} \\
% }\addtocounter{footnote}{-1}\endgroup

% %%% VLDB block end %%%


% \ifdefempty{\vldbavailabilityurl}{}{
%   \vspace{.3cm}
%   \begingroup\small\noindent\raggedright\textbf{PVLDB Artifact Availability:}\\
%   The source code has been made available at \url{https://github.com/junyimei/flowwalker-artifact}.
%   \endgroup
% }


\section{Introduction} \label{sec:introduction}

Random walk (RW) is a practical approach to extract graph information and is widely used in real-world applications such as social network analysis~\cite{socialnetwork}, recommendation systems~\cite{recommendation}, and knowledge graphs~\cite{knowledgegraph}. Take the friend recommendation in Douyin (a popular social media developed by ByteDance) as an example. In the recommendation graph, vertices represent users, and edges depict diverse user interactions such as co-liking, co-favoring, etc.
RW is used to generate random walk sequences serving the Graph Neural Network (GNN)~\cite{gnn,zhu2019aligraph,tripathy2023distributed,agl} tasks for personalized friend recommendations.
However, the computational demands of RW are substantial.
For instance, on a recommendation graph snapshot with 227 million users and 2.71 billion edges, RW takes up to 3.5 hours, contributing to 35\% of the end-to-end training duration.
Since recommendation graphs are undergoing frequent updates, ensuring the prompt completion of the RW tasks becomes vital for maintaining service quality. Consequently, there is an urgent need to accelerate RW computations.

\begin{figure}[t]\small
    \centering
    \begin{subfigure}[t]{0.23\textwidth}
        \centering
        \includegraphics[width=\textwidth]{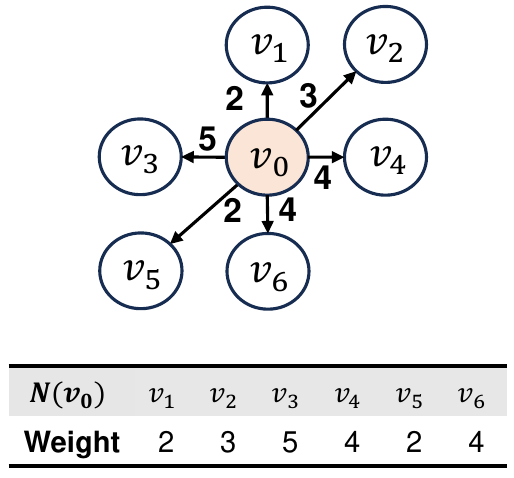}
        \caption{An example graph.}
        \label{fig:graph}
    \end{subfigure}
    \begin{subfigure}[t]{0.23\textwidth}
        \centering
        \includegraphics[width=\textwidth]{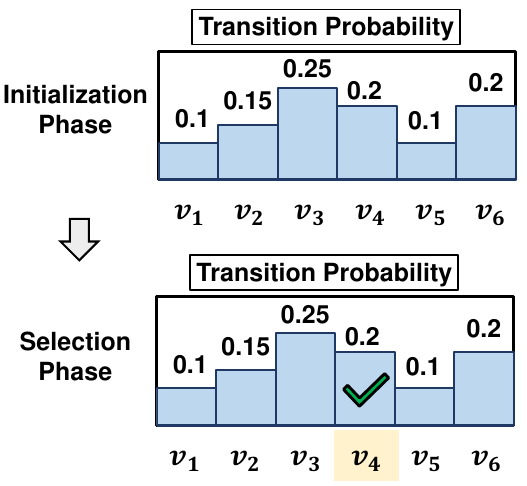}
        \caption{Sampling a neighbor of $v_0$.}
        \label{fig:rw}
    \end{subfigure}
    \caption{The procedure for sampling a neighbor of $v_0$.}
    \label{fig:paradigm}
\end{figure}

Recognizing the significance of the problem, researchers have conducted comprehensive studies~\cite{wang2020graphwalker,ThunderRW,knightking} to parallelize RW on multi-core CPUs.
Some works modify state-of-the-art graph processing frameworks to support RW algorithms, but they treat RW the same as traditional graph algorithms and ignore its
unique properties~\cite{dgl,graphchi,DrunkardMob}.
Thus specialized graph sampling frameworks have been proposed to maximize the overall sampling throughput.
For instance, GraphWalker~\cite{wang2020graphwalker} introduces a partition-based method for out-of-core computation. ThunderRW~\cite{ThunderRW} optimizes cache utilization to enhance in-memory computation.
These frameworks work well in \emph{static graph random walk} (SGRW) such as DeepWalk~\cite{deepwalk}, where the transition probability remains constant. Specifically, they execute SGRW in two phases: 1) a preprocessing phase that computes the transition probability table for each vertex, and 2) a computation phase that runs random walk queries.
As shown in Figure \ref{fig:paradigm}, this approach greatly diminishes the sampling cost~\cite{ThunderRW} by avoiding the initialization of the probability table at every step. However, this preprocessing strategy cannot process \emph{dynamic graph random walks} (DGRW), where transition probabilities are dynamically determined during runtime as in Node2Vec~\cite{node2vec} and MetaPath~\cite{metapath}. As a result, the computational complexity surges in DGRW as each step requires scanning the neighbors to calculate the transition probability table. For instance, ThunderRW can execute DeepWalk on the previously discussed recommendation graph in approximately 150 seconds with the preprocessing strategy; however, it exceeds an eight-hour time limit when running Node2Vec.

Recently, DGRW has gained popularity over SGRW due to its ability to capture temporal structure relations (i.e., the state of each query), rendering it a more powerful tool~\cite{tan2023lightrw,node2vec}. Researchers have turned to GPU acceleration to enhance DGRW performance leveraging their high-bandwidth on-board memory and massive computing power.
For example, C-SAW ~\cite{csaw} parallelizes \emph{inverse transform sampling}~\cite{its} on GPU, and Skywalker~\cite{skywalker} proposes a GPU-based \emph{alias table sampling}~\cite{alias} method. However, we uncover several fundamental limitations in existing GPU-based frameworks that lead to significant performance constraints.

First, these frameworks require extensive memory space to facilitate the query execution. They necessitate an $O(d)$ memory buffer to store the transition probability table for each query, where $d$ denotes the degree of the vertex that is being sampled. Since dynamic memory allocation can be costly, they opt to pre-allocate a buffer with $O(d_{max})$ size where $d_{max}$ denotes the maximum vertex degree in the graph. This approach can consume vast amounts of memory, especially when dealing with real-world graphs characterized by significant skewness.
In the case where $d_{max}$ in \emph{twitter} reaches $3 \times 10^6$, a buffer size of around 11.45 MB is required for every single query.
Though GPUs offer powerful computing capabilities, the limited memory space restricts concurrent parallelism (i.e., queries processed simultaneously) and reduces available space for graph data.

Second, these frameworks disregard the load imbalance issue emanating from both workload and hardware characteristics. The workload at each step is governed by the vertex degree, and the degree skewness among vertices can lead to workload imbalance. Besides, despite that RW is embarrassingly parallel, the concurrent execution capability of modern GPUs, which can support tens of thousands of threads, exacerbates the load imbalance problems among computing resources.
While C-SAW overlooks these concerns, Skywalker handles sampling tasks of varying degrees with warps or blocks, which leads to burdensome memory costs as well as communication overhead.

In this paper, we introduce \textbf{FlowWalker}, a GPU-based DGRW framework that performs fast sampling at minimal memory cost. We design a \emph{sampler-centric} computation model, which abstracts the computation from the hardware perspective. Under this model, an RW application is conceptualized as a set of discrete sampling tasks, where each task aims to randomly select a vertex from a specified vertex set. The GPU threads are systematically organized into a collection of samplers, which efficiently process these tasks. This abstraction narrows the RW computation down to two crucial problems: 1) devising efficient samplers; and 2) formulating an effective scheduling mechanism that assigns tasks to the samplers according to workload characteristics.

Inspired by sampling on streams, we design a parallel sampling method based on the \emph{reservoir sampling} technique~\cite{vitter1985reservoir}. This method is tailored for GPU optimization and is sufficiently adapted to handle vertices with varied degrees. Our design significantly reduces the space complexity of handling a sampling task from $O(d)$ to $O(1)$, thereby facilitating the concurrent execution of a substantial number of tasks. Coupled with efficient samplers, we develop a high-performance processing engine based on a multi-level task pool that distributes tasks among the samplers. Benefiting from its sampler design and processing engine, FlowWalker attains notable memory efficiency, with no auxiliary data structures in the global memory to streamline computation. Thereby, FlowWalker effectively tackles the challenges of limited query concurrency and load imbalance, optimizing the utilization of computational resources.

We showcase the generality of FlowWalker by implementing four representative algorithms, including DeepWalk~\cite{deepwalk}, PPR~\cite{ppr}, Node2Vec~\cite{node2vec}, and MetaPath~\cite{metapath}. We compare performance against ThunderRW~\cite{ThunderRW}, the state-of-the-art CPU-based framework; Skywalker~\cite{skywalker}, a GPU-based approach; DGL~\cite{dgl}, the widely used GNN framework; and LightRW~\cite{tan2023lightrw}, the state-of-the-art FPGA-bas\-ed approach. We conduct extensive experiments on ten real-world graphs, the size of which scale from millions to billions. Experiment results show that 1) FlowWalker stands as the sole GPU-based solution that able to support all of the four algorithms above; 2) FlowWalker consistently completes all test cases within a time frame of 2.2 hours, achieving up to 752.2$\times$ speedup over competitors, whereas DGL, LightRW, ThunderRW and Skywalker either exceed an eight-hour limit or encounter memory constraints; and 3) FlowWalker has negligible memory cost by getting rid of auxiliary data structures. In summary, we make the following contributions in this paper:

\begin{itemize}
    \item We introduce FlowWalker, a memory-efficient and high-performance GPU-based random walk framework, which leverages a sampler-centric computation model.
    \item We propose an efficient parallel sampling method for GPU based on reservoir sampling. This method greatly diminishes the space complexity, thereby substantially accelerating the sampling process.
    \item We develop a concise scheduling mechanism to efficiently channel a vast number of fine-grained tasks through samplers of different granularities. This mechanism enhances overall efficiency and adaptability.
\end{itemize}

\textbf{Paper Organization.} Section 2 introduces backgrounds. Section 3 gives an overview of the system. Sections 4 and 5 elaborate on the sampling method and computation engine, respectively. Section 6 details our experiment as well as case study. Section 7 concludes this paper.

\section{Background} \label{sec:background}

We introduce the preliminary and the background related to our work in this section.

\subsection{Graph Random Walk}

Let $G=(V, E)$ denote a directed graph where $V$ is the set of vertices and $E$ is the set of edges. Given a vertex $v \in V$, $N(v)$ is the neighbor set of $v$ and $d(v)$ is the degree, i.e., $|N(v)|$. Given an edge $e(u, v) \in E$, $w(u, v)$ and $l(u, v)$ represent its weight and label respectively.

Algorithm \ref{algo:common_paradigm} presents a common RW computation paradigm. An RW algorithm has a set $\mathbb{Q}$ of random walk queries. A query $Q$ begins at a start vertex. At each step, $Q$ randomly selects a neighbor $u$ of the current residing vertex $Q.cur$ and moves to it.  The operation is performed in two phases: 1) the \emph{initialization phase} calculates the \emph{transition probability} $p(u)$ for each neighbor $u$; and 2) the \emph{selection phase} randomly picks a neighbor given the distribution. $Q$ records the walk sequence in $Q.seq$ and stops until meets a specified condition, for example, $Q.seq$ reaches a length threshold. The outputs are the query sequences. Assume that the current residing vertex is $Q.cur = v$. The selection of a neighbor involves sampling $u$ from $N(v)$ based on a transition probability $p(u)$, determined by a weight function $f$ applied to the edge $e(v, u)$. For instance, if we define $f(e(v, u)) = w(v, u)$, then $p(u) = \frac{w(v, u)}{\sum_{u' \in N(v)} w(v, u')}$, which is a normalized value. To simplify the presentation, we refer to the transition probability $p(u)$ as the relative chance (e.g., the edge weight $w(v, u)$) of $u$ being selected without normalization in the subsequent discussions.

\setlength{\textfloatsep}{0pt}
\begin{algorithm}[t]
	\caption{Random Walk Computation Paradigm}
	\label{algo:common_paradigm}
	\small
	\SetKwRepeat{Do}{do}{while}
	 \KwIn{a graph $G$ and a set $\mathbb{Q}$ of RW queries\;}
	 \KwOut{the sequence of each query $Q \in \mathbb{Q}$\;}
      
	 \For{$Q \in \mathbb{Q}$}{
	    \Do{Stop($Q$) is false}{
     \tcc{The initialization phase.}
	 \ForEach{$u \in N(Q.cur)$} {       
            Calculate $u$'s transition probability $p(u)$\;
	    }
     \tcc{The selection phase.}
        Select a $u \in N(Q.cur)$ given $p(u)$ and add it to $Q.seq$\;
     }
	 }
     \KwRet $\mathbb{Q}.seq$\;
\end{algorithm}

Graph random walk algorithms are broadly divided into two categories based on the transition probability property: \emph{static graph random walk} (SGRW) and \emph{dynamic graph random walk} (DGRW). In SGRW applications like DeepWalk and PPR, the transition probability is fixed throughout the computation. This allows for calculating values in a pre-processing stage (as discussed in Section \ref{sec:introduction}), which significantly reduces computational complexity by eliminating the initialization phase in Algorithm \ref{algo:common_paradigm}. 
In contrast, the transition probability of DGRW relies on the query states and requires determination during runtime. Consequently, the initialization is postponed to the computation step. 
Next, we will introduce two representative DGRW algorithms. 

\textbf{MetaPath}~\cite{metapath} is a widely used algorithm for representation learning in heterogeneous networks~\cite{dong2017metapath2vec}. Within MetaPath, an edge label schema $l_1 \rightarrow \ldots \rightarrow l_i \ldots \rightarrow l_k$ constrains the walk sequence $Q.seq$ of a random walk query. Specifically, the labels of adjacent vertices in the sequence must align with the schema, i.e., $l(Q.seq[i], Q.seq[i + 1]) = l_{i}$. Suppose the current residing vertex is $Q.cur = v$, where $v$ is the $i$-th vertex in $Q.seq$. The transition probability for selecting a neighbor $u \in N(v)$ is defined by Equation \ref{eq:metapath}. The weighted version of MetaPath incorporates the edge weight into the calculation by multiplying it with the transition probability $p(u)$.
\begin{equation} \label{eq:metapath}
p(u)=
\begin{cases} 
1, &\text{if}\ l(v, u) = l_{i},\\
0, &\text{otherwise}.\\
\end{cases}    
\end{equation}

\textbf{Node2Vec}~\cite{node2vec} is a second-order RW algorithm, where the transition probability is dependent on the last visited vertex. Assuming that $Q.cur$ is $v$, then the transition probability $p(u)$ for selecting a neighbor $u \in N(v)$ is governed by Equation \ref{eq:n2v}, in which $v'$ represents the last visited vertex before $v$ and $dist(v', u)$ denotes the distance between $v'$ and $u$. $a$ and $b$ are two hyperparameters that modulate the random walk behavior. Similar to MetaPath, the edge weight $w(v, u)$ can be factored into the computation by multiplying it with the computed transition probability $p(u)$.

\begin{equation}\label{eq:n2v}
\setlength{\abovedisplayskip}{0pt}
    \setlength{\belowdisplayskip}{0pt}
p(u)=
\begin{cases}
\frac{1}{a}, &\text{if}\ dist(v', u)=0,\\
1, &\text{if}\ dist(v', u)=1,\\
\frac{1}{b}, &\text{if}\ dist(v', u)=2,\\
0, &\text{otherwise}.\\
\end{cases}    
\end{equation}

In addition to Node2Vec and MetaPath, methods such as Hetespaceywalk \cite{HeteSpaceyWalk} exemplify the application of DGRW. Representation learning methods on Heterogeneous Information Networks (HINs) \cite{shi2016discriminative,lao2010relational} are typically grounded in DGRW, necessitating consideration of label information—akin to the MetaPath approach. DGRW is also used for similarity measurement \cite{wu2016remember,cosimrank} and community detection \cite{deng2017finding,boldi2012arc}. In ByteDance, there are massive graphs with vertex labels such as users, videos, and advertisement items. Taking the advertisement recommendation scenario in Douyin as an example, we need to generate random walk sequences for each user and advertisement item based on specific meta-paths, such as user-item-user. Subsequently, the embeddings are trained to serve as inputs for the recommendation models. The practical necessity for dynamic walk algorithms in real-world business scenarios has motivated us to commence work on FlowWorker.

\subsection{Sampling Methods}\label{sec:sampling}

In the context of our study, sampling is the process of selecting a vertex $u$ from a neighbor set $N(v)$ based on the transition probability distribution. Different frameworks implement this operation through various sampling methods. ThunderRW, for example, offers \emph{inverse transform sampling} (ITS)~\cite{its}, \emph{rejection sampling} (RJS)~\cite{rejection}, and \emph{alias table sampling} (ALS)~\cite{alias,hubschle2022parallelsampling}, allowing users to choose the method most suitable for the algorithm's property. C-SAW~\cite{csaw} and Skywalker~\cite{skywalker} utilize ITS and ALS, respectively. However, both methods require an $O(d)$-sized memory buffer to store the transition probability, which, as discussed in Section \ref{sec:introduction}, consumes substantial memory and can lead to significant performance issues. Contrastingly, RJS requires only $O(1)$ space to store the maximum transition probability, employing a “trial-and-error” selection approach. However, this method comes with its drawbacks: the non-deterministic running time of randomized selection is heavily affected by the underlying probability distribution, and the process leads to numerous random memory accesses. These factors make RJS challenging to implement efficiently on GPUs.

Contrary to other methods, \emph{reservoir sampling} (RS)~\cite{vitter1985reservoir,chao1982general} is tailored for sampling streaming data. As outlined in Algorithm \ref{algo:serial}, RS operates on a vertex sequence $S$ with length $n$. $W_P$ maintains the prefix sum of weights, and $selected$ stores the index of the vertex chosen from $S$. Upon encountering a vertex at position $i$, RS updates $W_P$ and generates a random number. If this number is smaller than the transition probability $\frac{W[i]}{W_P}$, RS updates the $selected$ index accordingly (Line 4). Ultimately, RS returns the last selected vertex. Notably, the space complexity of RS is $O(1)$, and the time complexity is $O(d)$ given a neighbor set $N(v)$ with $d$ vertices as the input. While both ITS and ALS require only a single random number, RS necessitates generating a random number for each element. Although this might pose a challenge for CPUs, it is well-suited for GPUs, which offer ample computational resources.

\setlength{\textfloatsep}{0pt}
\begin{algorithm}[t]
\caption{Sequential Weighted Reservoir Sampling}
\label{algo:serial}
\small
\SetKwRepeat{Do}{do}{while}
\KwIn{a vertex sequence $S$, the corresponding weight sequence $W$, the sequence length $n$\;}
\KwOut{a vertex sampled from $S$ based on $W$\;}

$ W_P \leftarrow 0, selected \leftarrow 0$\;

\For{$i\leftarrow1\ to\ n$}{
    $W_P\leftarrow W_P + W[i]$\;
    \lIf{\textsc{random($0, 1$)}$< \frac{W[i]}{W_P}$}{$selected\leftarrow i$}
}  
\KwRet $S[selected]$\;
\end{algorithm}

\subsection{GPU-based Random Walk Frameworks}\label{sec:gpu_framework}

Researchers have proposed several works to accelerate RW applications using GPUs. NextDoor~\cite{nextdoor} is a graph sampling framework utilizing the RJS sampling method. It adopts the offline computation mode, which calculates the maximum weight for a neighbor set during the pre-processing stage. When executing random walk queries, NextDoor only performs the selection phase of the sampling. Therefore, NextDoor cannot support variant DGRW applications. Note that NextDoor implements unweighted Node2Vec by choosing the maximum value from $(1, \frac{1}{a}, \frac{1}{b})$ to bypass the initialization phase. The implementation cannot be generalized to weight\-ed Node2Vec and other DGRW applications such as weighted MetaPath. During runtime, NextDoor follows the BSP~\cite{bsp} model, advancing all queries by a single step at a time. NextDoor, which can sample multiple vertices from a neighbor set, ensures load balance by allocating threads according to the number of sampling results.

Distinct from NextDoor~\cite{nextdoor}, C-SAW~\cite{csaw} supports DGRW and employs the ITS sampling method. It adopts a query-centric computation model, assigning each query to a warp and executing them synchronously in a step-by-step fashion using the BSP~\cite{bsp} model. Although C-SAW optimizes ITS for GPUs to speed up computations, it falls short in supporting queries with variable walk lengths, such as PPR, due to its synchronized execution approach.

Skywalker~\cite{skywalker,wangl2023skywalker+} parallelizes the ALS sampling methods and optimizes memory access by compressing alias tables. To address the load imbalance caused by varying vertex degrees, Skywalker employs versatile samplers tailored to vertices with different degrees. To further mitigate load imbalance among thread blocks, it introduces a queue to distribute queries across blocks. As queries can have different lengths, the space complexity of the queue is $O(L_{max} \times |\mathbb{Q}|)$ where $L_{max}$ is the maximum length of queries.

Despite these advancements, both C-SAW and Skywalker possess foundational limitations, as discussed in Section \ref{sec:introduction}, which restrict their efficiency in handling large graphs. Besides, frameworks like GraSS \cite{grass} focus on graph compression. 
This technique is complementary to our work and can be integrated with FlowWalker to further minimize memory usage.

\subsection{Other Related Works}

Given the critical role of Random Walk (RW) applications, numerous studies have focused on optimizing CPU-based graph random walk frameworks. NosWalker~\cite{wang2023noswalker}, DrunkardMob~\cite{kyrola2013drunkardmob}, and GraphWalker~\cite{wang2020graphwalker} are designed to handle graphs that exceed available memory. ThunderRW~\cite{ThunderRW} optimizes in-memory computation by improving cache utilization. KnightKing~\cite{knightking} and FlashMob~\cite{yang2021random} are distributed frameworks that address communication and memory bandwidth utilization. Nevertheless, all these frameworks are optimized for SGRW, though some of them (e.g., ThunderRW) can execute DGRW. Additionally, research efforts have been made to optimize memory usage for random walks on both static and streaming graphs~\cite{shao2020memory,papadias2022space}.

Recently, Tan et al.~\cite{tan2023lightrw} introduce an FPGA-based approach to accelerate DGRW. They develop a parallel reservoir sampling meth\-od on FPGAs, akin to Algorithm \ref{algo:direct_parallel_rs}. Despite the similarities, the fundamental differences in the underlying hardware architectures set our approaches apart. LightRW's emphasis lies in customizing hardware to optimize pipeline execution and memory access during sampling. In contrast, GPU architectures are fixed, with threads grouped into thread blocks at runtime. Our approach involves a meticulous exploration of the design space to adapt reservoir sampling to the unique demands of GPU workloads and hardware characteristics. The inherent distinctions in hardware architectures influence our respective sampling algorithms, system designs, and research focuses.

\section{An Overview of FlowWalker} \label{sec:overview}

\begin{figure}[t]
  \centering
  \includegraphics[width=\linewidth]{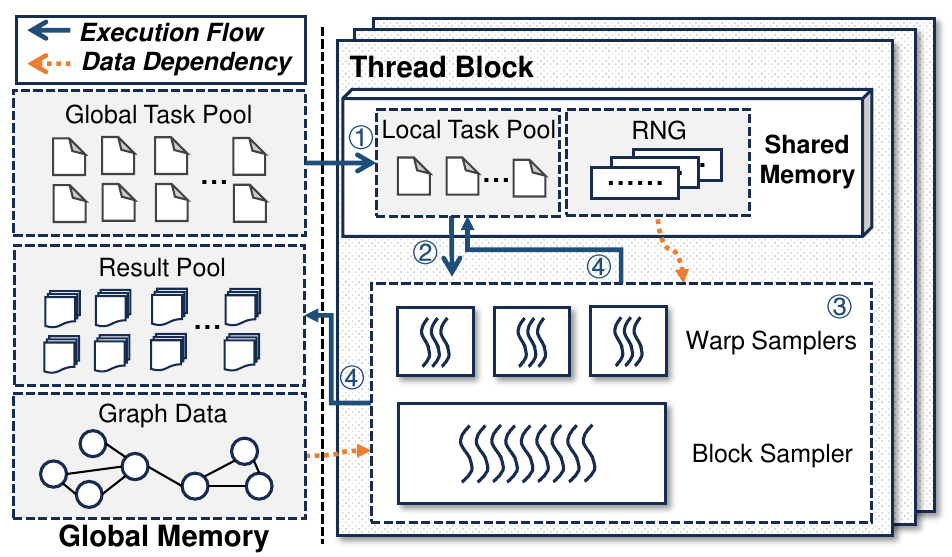}
  \caption{System Design Overview of FlowWalker. The execution flow is organized as follows: \textcircled{1} Thread blocks fetch tasks from the global task pool into its local task pool. \textcircled{2} Tasks are dispatched to the appropriate sampler based on the vertex degree. \textcircled{3} Warp and block samplers execute the sampling tasks. The process necessitates graph data stored in the global memory and random number generators (RNG) stored in the shared memory. \textcircled{4} The query states in the local task pool are updated and the sampling results are recorded.}
  \label{fig:system}
\end{figure}

\textbf{Computation Model.} Different from the query-centric model, we propose the \emph{sampler-centric} model that abstracts the computation from the hardware perspective. Specifically, an RW application consists of massive random walk queries each of which is a sequence of steps. A step performs one sampling operation, which selects a neighbor from the neighbor set of the current residing vertex and updates the query. Therefore, an RW application can be viewed as a set of sampling tasks. The computation on GPUs is to organize threads to a set of samplers to perform these sampling tasks efficiently until all queries are complete. 

\noindent\textbf{System Design.} Based on the sampler-centric model, we design FlowWalker, a memory-efficient and high-performance GPU-based DGRW framework. We propose a parallel reservoir sampling meth\-od that can perform the sampling with $O(1)$ memory cost. Besides, an efficient computation engine is implemented to guide global task scheduling and computation inside a thread block. 

Figure \ref{fig:system} gives an overview of our system design. In Flow\-Walker, a thread block is an independent worker whose threads are organized into samplers with different parallelism. Particularly, given a set of sampling tasks, a thread block adopts a two-stage execution scheme to handle variant workloads among these tasks. In the first stage, threads are organized into warp samplers (i.e., a warp works as a sampler) to process small tasks. In the second stage, all threads in the same thread block form a block sampler (i.e., a block works as a sampler) to handle large tasks. A multi-level task pool based dynamic scheduling mechanism is adopted to keep load balance among computing resources. A thread block has a local task pool that maintains the queries assigned to it. Once a query is completed, it will fetch a new query from the global task pool. The fine-grained scheduling method requires no communication and synchronization among blocks and achieves good load balance. Additionally, it gets rid of auxiliary data structures in the global memory, and a small amount of intermediate data can be held inside the shared memory, which is a type of fast-speed GPU memory. In terms of APIs, our framework adheres to the conventions established by prior works~\cite{ThunderRW, csaw, skywalker, wang2020graphwalker}. Therefore, we omit the details for brevity.

Benefiting from the designs mentioned above, FlowWalker is able to perform memory-efficient sampling with no data structures stored in the global memory to assist the execution. This significantly benefits GPU-based RW because GPUs have abundant computing resources but limited memory space. We will introduce the sampling method and the engine in Sections \ref{sec:sampler} and \ref{sec:engine}, respectively.

\section{Sampling Method} \label{sec:sampler}

Under the sampler-centric abstraction, sampling is the key operation in RW applications. As discussed in Section \ref{sec:background}, existing methods~\cite{its,rejection,alias} have severe performance issues on GPUs due to the large memory consumption of the intermediate data. Inspired by stream processing, we model the problem of choosing a neighbor as that of sampling an element from a stream. Therefore, we can adopt reservoir sampling (RS) to reveal the memory consumption issue because RS does not maintain a state for each element.

\begin{figure*}[t]
\setlength{\abovecaptionskip}{0pt}
    \setlength{\belowcaptionskip}{0pt}
    \captionsetup[subfigure]{aboveskip=0pt,belowskip=0pt}
    \centering  
    \includegraphics[width=\linewidth]{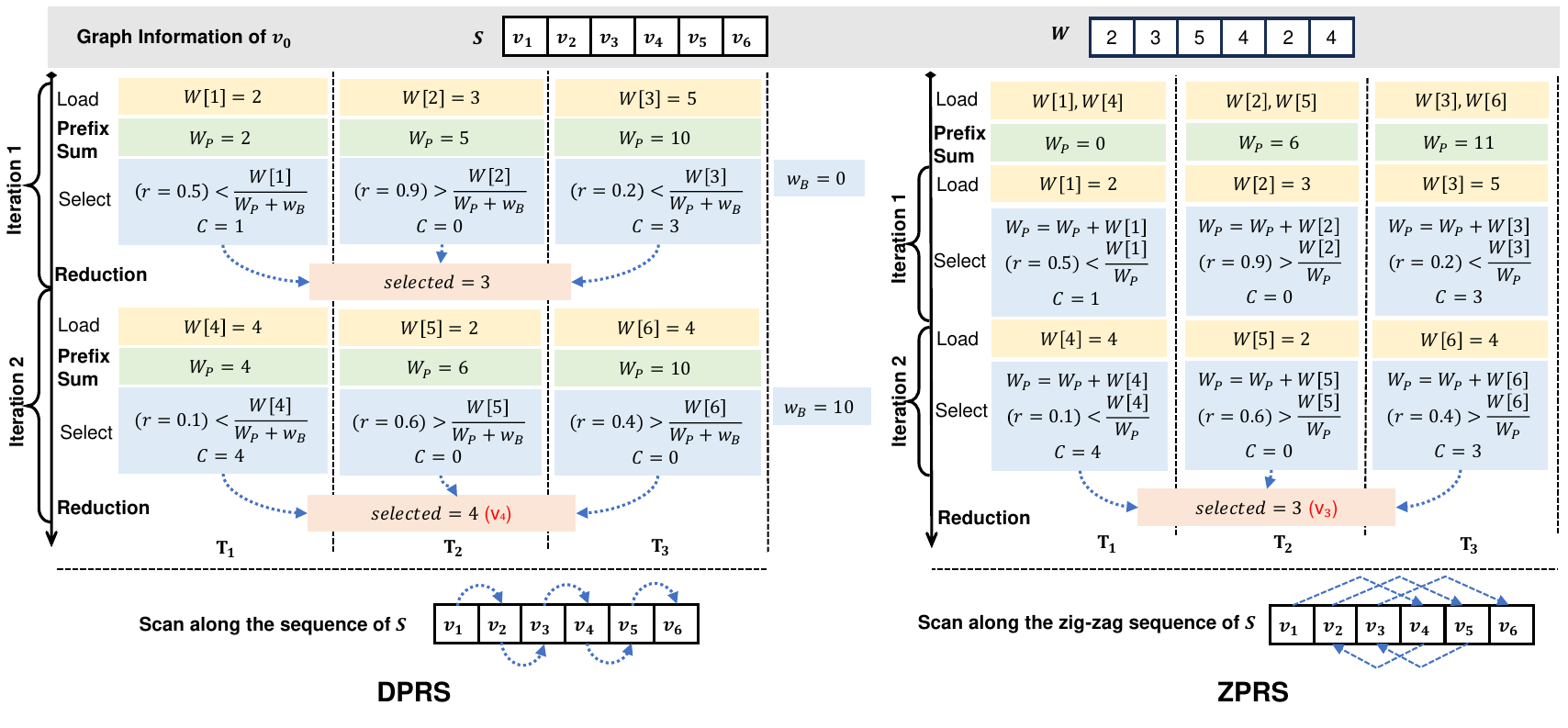}  
    \caption{The comparison of DPRS and ZPRS on sampling a neighbor of $v_0$ in Figure \ref{fig:graph} using three threads. DPRS scans $W$ once, but the number of collective operations depends on the number of iterations. ZPRS performs two collective operations only, but scans $W$ twice. Logically, DPRS scans along the sequence of $S$, whereas ZPRS scans in a zig-zag order of $S$.}   
    \label{fig:sampler_example}   
\end{figure*}

\subsection{Direct Parallel Reservoir Sampling}

Given a sequence $S$ of $n$ vertices, the corresponding weights $W$ and a group of $k$ threads (e.g., a warp), our goal is to parallel reservoir sampling which selects a vertex $v$ from $S$ based on $W$. Moreover, we want to keep $k$ threads having coalesced memory access patterns to fully utilize GPUs. Recall that reservoir sampling scans $S$ along the sequence order with the probability of replacing the selected vertex with $v_i$ as $\frac{w_v}{W_i}$ where $v_i$ is the $i$th vertex in $S$, $w_v$ is the weight of $v$, and $W_i = \sum_{j = 1} ^ {i} W[j]$ (i.e., the sum of weights of vertices before $v_i$ in $S$). If $v_i$ is picked, then we replace the selected vertex with $v_i$. Reservoir sampling returns the last selected vertex as the sampling result.

A straightforward idea of parallelization is to sample a vertex from $k$ consecutive vertices in parallel in each iteration and repeat until all vertices are processed. We call this method the \emph{direct parallel reservoir sampling} (DPRS) algorithm. Algorithm \ref{algo:direct_parallel_rs} depicts the details. In a certain iteration (Lines 4-11), we first read weights from $W$ for $k$ vertices in parallel with thread $j$ holding value $W_L[j]$. Next, we compute the prefix sum $W_P$ for the $k$ values in parallel. $w_B$ maintains the sum of weights in previous iterations, i.e. vertices from $S[1]$ to $S[i \times k]$. Therefore, thread $j$ selects the vertex $S[j + i \times k]$ with the probability $\frac{W_L[j]}{W_P[j] + w_B}$ (Line 9). We then set the selected index to the maximum value in $C$ (i.e., the maximum sequence index selected by these $k$ threads) and update $w_B$ (Lines 10-11). Finally, we return the sampled vertex given the index (Line 12). Note that returning $S[0]$ denotes that no vertex is selected, for example, no label can match the constraint in MetaPath.

\begin{example}
    Figure \ref{fig:sampler_example} presents a running example of DPRS where $n=6$ and $k=3$. At Iteration 1, threads $T_{1-3}$ first load weights of $v_{1-3}$ in parallel and then compute their prefix sum. After that, they perform the selection independently. For example, $T_1$ sets the selected index $C$ to $1$ since the random number value $r=0.5$ is less than $\frac{W_L}{W_P + w_B} = 1.0$. At the end of the iteration, DPRS performs a parallel reduction to get the last selected index (i.e., the maximum $C$ among $T_{1-3}$), which is the selected item at this iteration. Additionally, DPRS sets $w_B$ to $10$, which is the $W_P$ value held by $T_3$. DPRS continues its computations until all elements have been processed. The result is 4 and the selected vertex is $v_4$. The parallel sampling order is equivalent to the order of $S$.
\end{example}

\textbf{Analysis.} Given the vertex $v = S[j + i\times k]$, thread $j$ updates the selected vertex with the probability of $\frac{w_v}{\sum_{m = 1} ^ {j + i \times k} W[m]}$. The max operation keeps the algorithm to return the last picked vertex. Therefore, Algorithm \ref{algo:direct_parallel_rs} intuitively has the same logic as Algorithm \ref{algo:serial} though it runs in parallel, and Proposition \ref{prop:drs_correctness} holds.

\begin{proposition} \label{prop:drs_correctness}
    Given a sequence $S$ of vertices and the corresponding weight sequence $W$, Algorithm \ref{algo:direct_parallel_rs} picks $v$ with the probability $\frac{w_v}{\sum W}$ where $w_v$ is the weight of $v$.
\end{proposition}

Next, we analyze the time cost of Algorithm \ref{algo:direct_parallel_rs}. Suppose that the cost of obtaining $W[i]$ is $\alpha$, that of communication among threads is $\beta$, and that of random number generation is $\gamma$. In Algorithm \ref{algo:direct_parallel_rs}, Line 6 accesses global memory, Lines 7 and 10 perform the parallel collective operations among $k$ threads, and Line 9 computes a random number in each thread. Therefore, the cost at one iteration is $\alpha + 2 \times \beta\log k + \gamma$. The time cost of the algorithm is $\ceil{\frac{n}{k}} \times (\alpha + 2 \times \beta \log k + \gamma)$. The time complexity is $O(\frac{n}{k} \times \log k)$, and the speedup over Algorithm \ref{algo:serial} is $O(\frac{k}{\log k})$.

Finally, we discuss the space complexity of Algorithm \ref{algo:direct_parallel_rs}. In addition to storing $S$ and $W$, we do not maintain a state for each vertex, while each thread only requires several local variables ($W_P$, $C$, and $w_B$, etc.). Therefore, the space complexity of the algorithm is $O(k)$ and that for one thread is $O(1)$.

\setlength{\textfloatsep}{0pt}
\begin{algorithm}[t]
	\caption{Direct Parallel Reservoir Sampling(DPRS)}
	\label{algo:direct_parallel_rs}
	\small
	\SetKwRepeat{Do}{do}{while}
        \SetKwFor{ParallelFor}{parallel for}{do}{}
	 \KwIn{a vertex sequence $S$, the corresponding weight sequence $W$, the sequence length $n$ and $k$ threads\;}
	 \KwOut{a vertex sampled from $S$ based on $W$\;}
      \ParallelFor{$j \leftarrow 1$ to $k$}{
          $C[j] \leftarrow 0, W_L[j] \leftarrow 0, W_{P}[j] \leftarrow 0$\;
      }
      $w_{B} \leftarrow 0$\;
      \For{$i \leftarrow 0 \text{ to } \lceil \frac{n}{k} \rceil - 1$}{
      \ParallelFor{$j \leftarrow 1$ to $k$}{
          $W_L[j] \leftarrow W[j + i \times k]$\;
       }
       $W_{P}\leftarrow$\textsc{parallel\_inclusive\_prefix\_sum($W_L, k$)}\;

       \ParallelFor{$j \leftarrow 1$ to $k$}{
          \lIf{\textsc{random($0, 1$)} $< \frac{W_L[j]}{W_{P}[j] + w_{B}}$}{$C[j] \leftarrow j + i \times k$}
       }
       \tcc{Get the maximum value in $C$.}
       $selected \leftarrow $\textsc{parallel\_reduction}($C, k$)\;
       $w_{B} \leftarrow w_B + W_{P}[k]$\;     
      } 
	 \KwRet $S[selected]$\;
\end{algorithm}

\subsection{Zig-Zag Parallel Reservoir Sampling} \label{sec:zprs_analysis}

Although DPRS accesses the global memory in a coalesced pattern, we find that DPRS can have performance issues when processing long vertex sequences. Specifically, a GPU thread group has a limited number of threads, for example, a warp has 32 threads. Consequently, given a long vertex sequence (e.g., millions of vertices), DPRS frequently performs parallel collective operations that incur expensive costs due to communication overhead among threads. As real-world graphs have vertices with large degrees and processing these vertices dominates the random walk cost, the performance issue degrades the computation speed.

To solve the problem, we design the \emph{zig-zag parallel reservoir sampling} (ZPRS), which not only has coalesced memory access patterns but also reduces the number of parallel collective operations. In particular, different from DPRS scanning and sampling vertices along the order of $S$, ZPRS scans vertices along the order but samples in a zig-zag order $S'$. Algorithm \ref{algo:zigzag_parallel_rs} presents the details. $S$ can be divided into $k$ sets where $S_j = \{v_m \in S| m \mod k = j\}$. We first compute the weight sum for vertices in $S_j$ and store the value to $W_L[j]$ (Lines 3-5). Next, we compute the exclusive prefix sum on $W_L$ such that $W_P[j] = \sum_{m = 1} ^ {j - 1} \sum_{v \in S_m} w_v$. After that, thread $j$ replaces the selected vertex with $v$ in the probability $\frac{w_v}{W_P[j]}$. To pick the last sampled vertex, we select the last item that is greater than 0 in $C$ in parallel (Line 11).

\begin{example}
    Figure \ref{fig:sampler_example} presents a running example of ZPRS where $n=6$ and $k=3$. Threads $T_{1-3}$ first load six weights in parallel at two iterations and then calculate the exclusive prefix sum. As this procedure is simple, we omit the details of the two iterations and directly show $W_P$ values. After that, $T_{1-3}$ performs the sampling independently. For example, at Iteration 1, $T_3$ first loads $W[3]$ and then sets the selected index $C$ to $3$ because the random number $r = 0.2$ is less than $\frac{W[3]}{W_P} = 0.31$. After processing all elements, $T_{1-3}$ performs a parallel reduction to get the last $C$ value such that $C > 0$. The result is $3$ and the selected item is $v_3$. The parallel sampling order is equivalent to along a zig-zag order of $S$.
\end{example}

\setlength{\textfloatsep}{0pt}
\begin{algorithm}[t]
	\caption{Zig-Zag Parallel Reservoir Sampling(ZPRS)}
	\label{algo:zigzag_parallel_rs}
	\small
	\SetKwRepeat{Do}{do}{while}
        \SetKwFor{ParallelFor}{parallel for}{do}{}
	 \KwIn{a vertex sequence $S$, the corresponding weight sequence $W$, the sequence length $n$ and $k$ threads\;}
	 \KwOut{a vertex sampled from $S$ based on $W$\;}
      \ParallelFor{$j \leftarrow 1$ to $k$}{
          $C[j] \leftarrow 0, W_L[j] \leftarrow 0, W_{P}[j] \leftarrow 0$\;
      }
      \For{$i \leftarrow 0 \text{ to } \lceil \frac{n}{k} \rceil - 1$}{
       \ParallelFor{$j \leftarrow 1$ to $k$}{
          $W_L[j] \leftarrow W_L[j] + W[j + i \times k]$\;
        }
       }
       $W_{P}\leftarrow$\textsc{parallel\_exclusive\_prefix\_sum($W_L, k$)}\;
       \For{$i \leftarrow 0 \text{ to } \lceil \frac{n}{k} \rceil - 1$}{
       \ParallelFor{$j \leftarrow 1$ to $k$}{
          $W_{P}[j] \leftarrow W_{P}[j] + W[j + i \times k]$\;
       \lIf{\textsc{random($0, 1$)} $< \frac{W[j + i \times k]}{W_{P}[j]}$}{$C[j] \leftarrow j + i \times k$}
        }
       }
       \tcc{Get the last item greater than 0 in $C$.}
       $selected \leftarrow $\textsc{parallel\_reduction}($C, k$)\;
	 \KwRet $S[selected]$\;
\end{algorithm}

\textbf{Analysis.} First, we prove Proposition \ref{prop:zrs_correctness} based on the correctness of Algorithm \ref{algo:serial}, which is proved in the technical report.

\begin{proposition} \label{prop:zrs_correctness}
    Given a sequence $S$ of vertices and the corresponding weight sequence $W$, Algorithm \ref{algo:zigzag_parallel_rs} picks $v$ with the probability $\frac{w_v}{\sum W}$ where $w_v$ is the weight of $v$.
\end{proposition}

\begin{proof}
    Consider a sequence $S$ of $n$ elements with corresponding weights $W$ and $k$ threads. Define $S_i$ as a sub-sequence of $S$ such that $S[j] \in S_i$ if $j \mod k = i$ for $1 \leqslant j \leqslant n$, and set $S_k = S_0$. This construction yields a new sequence $S' = (S_1, S_2, \ldots, S_k)$ and its associated weight sequence $W'$. As shown in Lines 7-10, each thread $i$ processes $S_i$ independently. Given 
    $v = S_i[j]$
    , thread $i$ replaces its current selected vertex with a probability $\frac{W'[j]}{\sum_{l = 1}^{j} W'[l]}$. Line 11 ensures that the element chosen by thread $i$ is replaced by the selection of thread $j$ if $i < j$. Consequently, parallel processing mirrors serial sampling along $S'$. By Proposition \ref{prop:drs_correctness}, each element $S'[i]$ is selected with probability $\frac{W'[i]}{\sum W'}$. So Proposition \ref{prop:zrs_correctness} holds.
\end{proof}

We next analyze the time cost of Algorithm \ref{algo:zigzag_parallel_rs}. Compared with DPRS, ZPRS only requires two collective operations (Lines 6 and 11). In contrast, ZPRS scans the weight sequence twice (Lines 3-5 and 7-10). Therefore, the time cost of ZPRS is $\ceil{\frac{n}{k}}\times (2 \times \alpha + \gamma) + 2 \times \beta \log k$. The time complexity is $O(\frac{n}{k} + \log k)$ and the speedup over the sequential method is $O(k \times (1 - \frac{k \log k}{n + k \log k}))$. 
When processing long sequences, ZPRS has a better speedup than DPRS and generally runs much faster than DPRS in practice, because modern GPUs have a big bandwidth and a large cache, e.g., A100 has 1.5-2 TB/s bandwidth and 40 MB L2 cache. But for the cases where the transition probability requires an expensive computation (i.e., $\alpha$ is high), ZPRS can run slower than DPRS in practice because it has to calculate the probability for each element twice. 
Experiment results in Section \ref{sec:detail_eval} confirm our analysis. The space complexity of ZPRS is $O(k)$, which is the same as DPRS.

\subsection{Implementation}

Both DPRS and ZPRS access global memory in a coalesced pattern. In their implementation, we focus on reducing the cost of collective operations $\beta$ and that of random number generation.
 
In principle, both DPRS and ZPRS can be executed in parallel with any number of threads. However, in practice, modern GPUs manage threads with warps, blocks, and grids. Moreover, they only support efficient communication and synchronization for warps and blocks. Due to this constraint, we implement the warp and block samplers, which execute with one warp and one block, respectively. The parallel collective operations have been extensively studied~\cite{kogge1973scan,hillis1986scan,sengupta2006scan,martin2012reduce}. In our implementation, we use CUB~\cite{cub} to conduct the prefix sum and reduction operations. Variables such as $C, W_L$, and $W_P$ can be held with a register, and the collective calculation merely requires a shared memory buffer.

The cuRAND library~\cite{curand} generates a random number by updating a \texttt{curandState}, which is a C struct containing a small integer array to record the generator state. As both DPRS and ZPRS generate a random number for each vertex in $S$, a simple method is to maintain an array of \texttt{curandState} for the warp (or block) with each thread having one state. However, this leads to uncoalesced global memory accesses. To resolve the issue, we transform the array of structures into a structure of arrays to optimize the memory access pattern. Similar to NextDoor~\cite{nextdoor}, we store this structure in shared memory to further accelerate the computation. The optimization can bring up to $20.3\times$ speedup in our experiment in Section \ref{sec:detail_eval}. Investigating the efficient generation of massive random numbers (e.g., each thread has a random number generator) on GPUs constitutes a compelling topic for future study.

\section{FlowWalker Engine} \label{sec:engine}

An RW application consists of massive random walk queries and each query is a sequence of walking steps. Steps from different queries can be processed independently, while steps from the same query have dependency. Under the sampler-centric computation model, threads in GPUs are organized into samplers and each step is a task unit. Specifically, given a step of a query, a sampler updates the query by selecting a neighbor of the current residing vertex. To process these tasks efficiently, we encounter two challenges caused by the workload and hardware properties. First, the workload of a step is determined by the degree of the current residing vertex. Due to degree skewness among vertices, workloads among different tasks are imbalanced. Second, although an RW application is embarrassingly parallel, modern GPUs support tens of thousands of threads executing concurrently, which leads to load imbalance issues among computing resources. Additionally, the communication and synchronization cost on GPUs is expensive.

In this section, we design an efficient walking engine on the top of our parallel reservoir samplers. In this engine, thread blocks are independent workers. Given a set of tasks, a thread block processes them by organizing its threads into different-level samplers (i.e., samplers with different threads) to handle variant workloads. Moreover, we design an effective scheduling mechanism based on multi-level task pools to keep load balance among workers. In the following, we will introduce the computation in a thread block, and then we will elaborate on the scheduling mechanism. Finally, the time and memory cost will be discussed.

\subsection{Computation} \label{sec:computation}

To address workload imbalance, we can organize thread blocks to warp and block samplers and assign tasks to different thread blocks based on their degrees. However, 
under the query-centric model, a query needs to move between different thread blocks frequently. As the communication and synchronization cost among blocks is very expensive in GPUs, this approach can incur significant overhead. Therefore, instead of moving queries among different blocks, FlowWalker sticks a query to a thread block and processes tasks with variant workloads.

\begin{figure}[t]
  \centering
  \includegraphics[width=\linewidth]{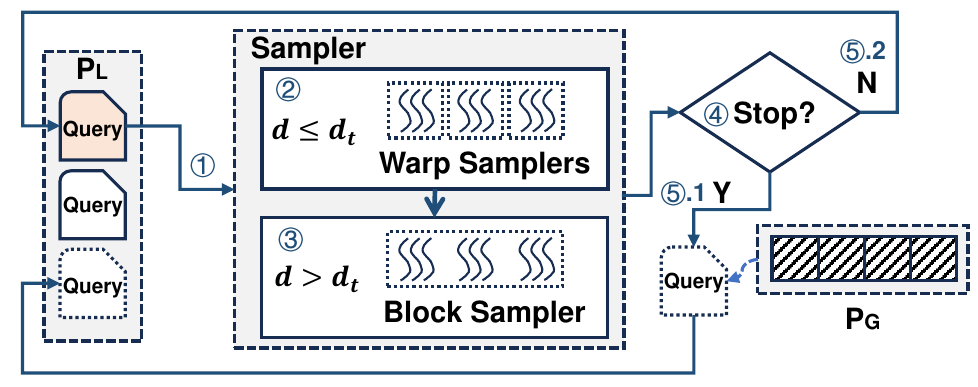}
  \caption{Computation in a thread block. A query will not be evicted from a thread block until stop conditions are met. Tasks are processed in two stages. First, warp samplers process tasks in which the degree of the current residing vertex is no greater than $d_t$. Then, the block sampler processes the remaining tasks. After sampling, if one query meets the stop conditions, a new query will be fetched from the global task pool ($P_G$) and added to the local task pool ($P_L$) (Step \textcircled{5}.1). Otherwise, we update the query state in $P_L$ (Step \textcircled{5}.2).}
  \label{fig:engine_local}
\end{figure}

Figure \ref{fig:engine_local} presents the computation in thread blocks. Each thread block has a \emph{local task pool} $P_L$ that maintains the queries assigned to it. An element in $P_L$ stores the status of a query $Q$, which has the current residing vertex $v$, the degree $d(v)$, the location of $N(v)$, the location of the result sequence $Q.seq$, and the length $|Q.seq|$ of the sequence. $P_L$ resides in shared memory because it is frequently accessed, while $N(v)$ and $Q.seq$ are stored in the global memory.

At the first stage, the thread block forms $\frac{|T|}{32}$ warp samplers to process the small tasks, the degrees $d(v)$ of which are no greater than a threshold $d_t$. $|T|$ denotes the number of threads in a block. As the warp is the basic scheduling unit in GPUs and executes independently, these samplers process small tasks in $P_L$ concurrently. Note that for the cases where the number of small tasks is less than warp samplers, the strategy still works well in modern GPUs because 1) the idle samplers incur a negligible cost, and 2) multiple thread blocks run concurrently on an SM to fully utilize hardware resources. After completing small tasks, the thread block forms a block sampler to process the remaining tasks one by one.

After the two stages, we store sampling results in the global memory and update the query status in $P_L$. If a query stops, we will get a new query from the global task pool, which will be introduced in the next subsection. In summary, queries in $P_L$ are processed iteratively and move one step at one iteration. A query will be processed in a specific block once it is fetched into the local task pool. This can eliminate the communication and synchronization costs among blocks. Moreover, the two-stage execution scheme processes tasks with variant workloads efficiently.

\subsection{Scheduling}

A simple method to handle massive queries is to evenly assign queries among workers (i.e., thread blocks). The static scheduling method works well on CPUs~\cite{ThunderRW,wang2020graphwalker}. However, we find that it can incur performance issues on modern GPUs because 1) GPUs have much higher parallelism than CPUs; and 2) thread block scheduling is transparent to users and certain thread blocks can start much later than others. To address this issue, we design a simple and effective dynamic scheduling method that cooperates with the two-stage computation scheme.

\begin{figure}[t]
  \centering
  \includegraphics[width=\linewidth]{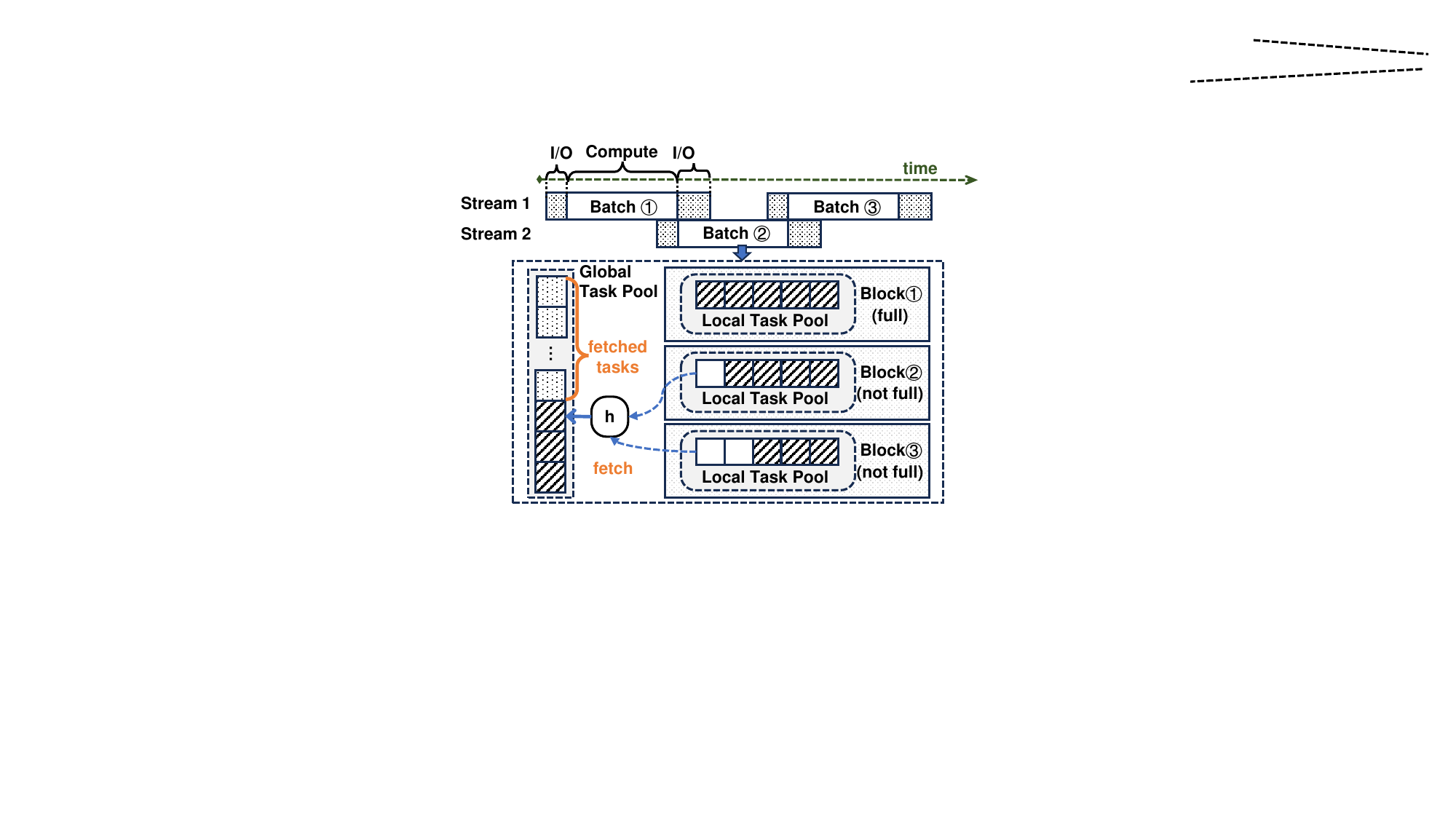}
  \caption{Queries are grouped into batches which execute alternatively in two CUDA streams. $h$ refers to the head pointer of the global task pool. Thread blocks fetch tasks in a preemptive way if they have empty slots.  }
  \label{fig:engine_global}
\end{figure}

Figure \ref{fig:engine_global} describes the dynamic scheduling strategy. We have a global task pool $P_G$, which keeps all queries in the device. Particularly, $P_G$ is an array where an element is the start vertex of a query. Correspondingly, the result pool is the array storing the query sequence with the size as $|P_G| \times L_{max}$ where $L_{max}$ is the maximum length of a query. The result sequence of a query is stored continuously. As discussed in Section \ref{sec:computation}, thread blocks execute independently. Upon finding that there are empty slots in the local task pool, they will fetch queries from the global task pool to fill these empty slots. The thread blocks fetch tasks from the head of $P_G$ in a preemptive manner. The concurrent accesses are supported by an atomic integer pointing to the first available queries in the pool. A thread block gets a query by increasing the integer atomically. The local task pool size is very small compared with the number of queries. Therefore, fine-grained scheduling can keep load balance among thread blocks to fully utilize computing resources. We do not adopt any work-stealing techniques because a query takes a short time and the communication and synchronization cost among thread blocks is expensive.

The number of queries residing on GPU is constrained by the result pool size. For the cases where the results exceed the result pool size, we process them in multiple batches. Specifically, we divide queries into multiple batches such that the result sequences of each batch can be held by the result pool. To overlap the GPU I/O time with computation time, we adopt the classical ping-pong buffer technique and process batches alternatively with two CUDA streams. The number of queries in a batch is determined by Equation \ref{eq:batch} where $M$ represents the total GPU memory size and $M_G$ is the memory allocated for the graph. $M_v$ is the memory required to store a single vertex. The overarching strategy aims to fully utilize available GPU memory for the result pool to minimize batch processing. Notice that: 1) the equation includes a division by two as a ping-pong buffer employs two alternating buffers; and 2) $L_{max} + 1$ includes the memory allocated for the start vertex for each query (i.e., the global task pool $P_G$). In summary, FlowWalker is capable of handling scenarios where the result sequence exceeds the available GPU memory.

\begin{equation} \label{eq:batch}
\setlength{\abovedisplayskip}{0pt}
    \setlength{\belowdisplayskip}{0pt}
|P_G| = \lfloor \frac{M - M_G}{2 \times (L_{max} + 1) \times M_v} \rfloor
\end{equation}

\subsection{Analysis and Comparison}

In the following, we analyze the cost of FlowWalker and compare it with C-SAW and Skywalker, two GPU-based systems.

\textbf{Memory Consumption.} The input is a graph $G$ and start vertices of queries $\mathbb{Q}$, and the output is the result sequence for each query. Their memory consumption is compulsory for all competing frameworks. Thus, we focus on the memory consumption for auxiliary data structures. The global task pool of FlowWalker is based on the array storing start vertices of walkers, which has no extra memory consumption, and the local task pool resides in the shared memory. Moreover, both warp and block samplers do not consume any global memory. Therefore, Flow\-Walker has no auxiliary data structures consuming the global memory.

In contrast, both C-SAW and Skywalker need an auxiliary data structure with $O(d_{max})$ to serve one query. This incurs expensive memory overhead for large graphs. Additionally, Skywalker uses a task pool with the memory consumption of $O(L_{max} \times |\mathbb{Q}|)$ to keep load balance among thread blocks. In summary, FlowWalker is memory-efficient, which brings two advantages: 1) FlowWalker can support larger graphs; and 2) the number of queries that can be processed simultaneously by FlowWalker is determined by computing resources, whereas that of C-SAW and Skywalker is limited by the available memory space. 

\textbf{Time.} We first compare the time cost of processing one step of a query. As analyzed in Section \ref{sec:sampler}, the time complexity of moving one step of a query using ZPRS is $O(\frac{d}{k} + \log k)$, while that of C-SAW is $O(\frac{d}{k}\times \log k + \log d)$ where $d$ is the degree of $Q.cur$. Skywalker uses the alias table sampling method to perform sampling. Although its time complexity is $O(\frac{d}{k} + \log k)$, the practical performance is slow due to the complex alias table building process. 

Next, we compare their techniques for keeping load balance. C-SAW can process a query with a warp only and uses a static scheduling method, which ignores both the load imbalance among tasks and thread blocks. Skywalker can adopt the parallelism based on degrees. However, Skywalker schedules queries among thread blocks with a global queue at each step. Consequently, each step requires a pop and a push operation, which incurs expensive overhead. And the queue consumes a large amount of memory space as discussed above. NextDoor assigns a single thread to a sampling function. 
This design ignores the variance of neighbor set sizes. Moreover, NextDoor operates in a BSP manner \cite{bsp}, advancing all queries by one step per iteration. This approach, however, may lead to two issues: 1) overhead from global synchronization, especially with queries of varying lengths such as PPR; and 2) the necessity to materialize all query results.

Under the sampler-centric model, FlowWalker handles variant tasks with different samplers and uses the multi-level task pool based scheduling strategy to keep load balance efficiently and effectively. Particularly, thread blocks can fetch a query by an atomic incremental operation, and a query sticks to the block until it is completed, which requires no communication and synchronization overhead among blocks. In our experiments, we show that Flow\-Walker runs much faster than its counterparts. Additionally, Flow\-Walker stands out as the only solution capable of handling cases where the result sequence exceeds available GPU memory.

\section{Experiments} \label{sec:experiments}

In this section, we conduct extensive experiments to evaluate the performance of FlowWalker.

\subsection{Experimental Setup}

We study five frameworks in the experiments. \textbf{DGL \footnote{https://github.com/junyimei/dgl}\cite{dgl}} is a widely used GNN framework. \textbf{LightRW\footnote{https://github.com/Xtra-Computing/LightRW} \cite{tan2023lightrw} (LRW)} is a FPGA-based DGRW framework. \textbf{ThunderRW\footnote{https://github.com/Xtra-Computing/ThunderRW} ~\cite{ThunderRW} (TRW)}, which is the state-of-the-art CPU-based framework, \textbf{Skywalker\footnote{https://github.com/wpybtw/Skywalker}~\cite{skywalker} (SW)}, which is a GPU-based framework, and \textbf{FlowWalker (FW)}, which is the GPU framework proposed in this paper. ThunderRW executes with the ITS sampling method, which achieves the optimal performance in the online computation mode. We also contemplated using \textit{C-SAW}\footnote{https://github.com/concept-inversion/C-SAW}. However, it encounters memory issues when handling more than $10 ^ 5$ queries. Therefore we exclude it from our experimental baselines. We do not involve NextDoor because it can only support the offline computation mode as discussed in Section \ref{sec:gpu_framework}.

\textbf{Implementation and Experiment Environments.}
\textit{FW} is implemented with $\sim$6000 lines of CUDA code. The experiments of \textit{DGL}, \textit{SW}, and \textit{FW} are conducted on a Linux server equipped with the 40 GB A100 GPU. It contains 108 streaming multiprocessors (SMs) each of which has 64 FP32 cores. The shared memory size of each SM is configured to 100 KB. The PCIe type is PCI-E 4.0 $\times$ 16, and the maximum bandwidth is 31.5GB/s. The server is equipped with one Intel(R) Xeon(R) Silver 4310 CPU and 256GB host RAM. We test \textit{TRW} on a Linux server equipped with one Intel Xeon Platinum 8336C CPU, which has 16 physical cores with hyper-threading enabled.  The size of the host RAM is 128 GB. \textit{LRW} is tested on HACC@NUS\footnote{https://xacchead.d2.comp.nus.edu.sg/} with an AMD Alveo U250 FPGA. We use NVCC of version 11.6, g++ of version 9.4.0 and the optimization flag \texttt{-O3} for compilation.

\begin{table}[t]
  \setlength{\abovecaptionskip}{0pt}
  \setlength{\belowcaptionskip}{0pt}
  \centering
  \caption{The detailed statistics of graphs. }
  \label{tab:data}
  \resizebox{\columnwidth}{!}{
    \begin{tabular}{
        cccccc
      }
      \toprule[0.8pt]

      {\bf Dataset} & \textbf{Name} & { $|V|$} & {\bf $|E|$} & {\bf $d_{max}$} & {\bf Size(GB)} \\
      \midrule[0.8pt]
      com-youtube   & YT            & 1.1 M    & 6 M         & 28K             & 0.05           \\
      \hline
      cit-patents   & CP            & 3.8 M    & 33 M        & 793             & 0.26           \\
      \hline
      Livejournal   & LJ            & 4.8 M    & 86 M        & 20K             & 0.66           \\
      \hline
      Orkut         & OK            & 3.1 M    & 234 M       & 33K             & 1.76           \\
      \hline
      EU-2015       & EU            & 11 M     & 522M        & 399K            & 3.93           \\
      \hline
      Arabic-2005   & AB            & 23 M     & 1.1B        & 576K            & 8.34           \\
      \hline
      UK-2005       & UK            & 39 M     & 1.6B        & 1.7M            & 11.82          \\
      \hline
      Twitter       & TW            & 42 M     & 2.4 B       & 3M              & 18.08          \\
      \hline
      Friendster    & FS            & 66 M     & 3.6 B       & 5K              & 27.16          \\
      \hline
      SK-2005       & SK            & 51 M     & 3.6 B       & 8.5M            & 27.16          \\
      \bottomrule[0.8pt]
    \end{tabular}
  }
\end{table}

\textbf{Datasets and Workloads.}
We select a variety of real-world graphs from different fields such as social networks, citations, and websites. The detailed statistics are listed in Table~\ref{tab:data}. YT, CP, LJ, OK, and FS are downloaded from Stanford SNAP~\cite{snapnets}, and EU, AB, UK, TW, and SK are from LAW~\cite{BoVWFI,BRSLLP}. We have data sizes ranging from tens of megabytes to tens of gigabytes (with weight). To keep consistent with previous work~\cite{ThunderRW,knightking}, we generate a real number randomly from an interval $[1,5)$ as the edge weight and an integer from the interval $[0,4]$ as the edge label.

We study DeepWalk, PPR, Node2Vec, and MetaPath in the experiments. For DeepWalk, we set the target depth to 80. For PPR, we set the stop probability to 0.2. For Node2Vec, we set the target length to 80, $a=2.0$ and $b=0.5$. For MetaPath, we set the schema to $(0,1,2,3,4)$. We issue a query from every vertex in the graph for DeepWalk, Node2Vec, and MetaPath. For PPR, $|V|$ queries start from the same vertex. We set the vertex to that with the maximum degree in $G$.
In detailed evaluation, we follow the settings of DeepWalk and set the number of queries to $10^6$ because \textit{SW} frequently encounters performance issues and has no valid experiment results for comparison.
For the comparison purpose, all applications, including SGRW are executed in the dynamic manner. As a result, the results on SGRW may diverge from those reported in previous papers \cite{ThunderRW,skywalker}, which are obtained with static mode.

\textit{FW} executes Node2Vec with DPRS, while the other three applications with ZPRS.
\textit{DGL} implements Node2Vec on CPUs, while the other three applications on GPUs. \textit{SW} does not support MetaPath because it cannot handle labeled graphs. \textit{LRW}, the FPGA-based framework, currently supports Node2Vec and MetaPath only. \textit{TRW} and \textit{FW} implement all these four applications.

\textbf{Metrics.}
The \emph{execution time} refers to the total time required for computation, excluding the time spent on loading the graph data into GPUs. The results are averaged through three runs. \textit{OOT} signifies that the method exceeds the time limit, which is set as 8 hours for our experiments, while \textit{OOM} indicates a memory overflow.
For a more comprehensive analysis, we employ \emph{NVIDIA Nsight Compute} to profile GPU \emph{memory consumption}.

\textbf{Parameters.} \textit{FW} requires two hyperparameters: the local task pool size $|P_L|$, and the degree threshold $d_t$. $|P_L|$ dictates the number of queries that a thread block can hold, and $d_t$ serves as the threshold for selecting between the warp sampler and block sampler. We empirically tune their values and set $|P_L|$ and $d_t$ to 64 and 1024, respectively, across our experiments. \textit{FW} achieves a good performance on the settings. Due to space limits, we include a detailed evaluation of hyperparameter impacts in the technical report.

\begin{table*}[t]
  \setlength{\abovecaptionskip}{0pt}
  \setlength{\belowcaptionskip}{0pt}
  \centering
  \caption{The overall comparison on execution time (seconds).}
  \label{tab:all}
  \begin{tabular}{c|c|cccccccccc}
    \bottomrule[0.8pt]
                                  & \bf Dataset & \bf YT   & \bf CP   & \bf LJ   & \bf OK   & \bf EU    & \bf AB     & \bf UK      & \bf TW      & \bf FS    & \bf SK      \\ \hline
    \multirow{4}{*}{\bf DeepWalk} & \bf DGL     & 0.93     & \bf0.30  & 1.25     & 1.84     & 68.11     & 3492.19    & OOM         & OOM         & OOM       & OOM         \\ \cline{2-12}
                                  & \bf TRW     & 6.90     & 3.81     & 14.28    & 20.86    & 739.97    & 3298.71    & OOT         & OOT         & 496.52    & OOT         \\ \cline{2-12}
                                  & \bf SW      & 7.82     & 3.20     & 21.89    & 28.88    & 431.61    & 1410.01    & OOT         & OOT         & OOM       & OOT         \\ \cline{2-12}
                                  & \bf FW      & \bf 0.45 & 0.42     & \bf 0.95 & \bf 0.99 & \bf 17.40 & \bf 59.86  & \bf 736.52  & \bf 2674.25 & \bf 24.26 & \bf 1509.83 \\ \hline
    \multirow{4}{*}{\bf PPR}      & \bf DGL     & 1.03     & 0.29     & 2.76     & 2.91     & 138.20    & 7728.80    & OOM         & OOM         & OOM       & OOM         \\ \cline{2-12}
                                  & \bf TRW     & 7.50     & 0.52     & 20.17    & 21.66    & 1900.78   & 3591.19    & OOT         & OOT         & 56.67     & OOT         \\ \cline{2-12}
                                  & \bf SW      & 4.10     & 0.85     & 10.85    & 11.70    & 690.55    & 1763.33    & OOT         & OOT         & OOM       & OOT         \\ \cline{2-12}
                                  & \bf FW      & \bf 0.23 & \bf 0.10 & \bf 0.74 & \bf 0.69 & \bf 32.60 & \bf 82.29  & \bf 1041.55 & \bf 897.61  & \bf 3.83  & \bf 2797.56 \\ \hline
    \multirow{5}{*}{\bf Node2Vec} & \bf DGL     & 273.71   & 132.65   & 428.92   & 583.50   & 15988.82  & OOT        & OOT         & OOM         & OOM       & OOM         \\ \cline{2-12}
                                  & \bf TRW     & 66.69    & 28.65    & 260.63   & 553.65   & 5936.80   & 23042.37   & OOT         & OOT         & 27329.18  & OOT         \\ \cline{2-12}
                                  & \bf SW      & 40.39    & 12.07    & 134.23   & 130.38   & 1065.75   & 2498.27    & OOT         & OOT         & OOM       & OOT         \\ \cline{2-12}
                                  & \bf LRW     & 12.68    & 7.13     & 18.16    & 24.70    & 758.57    & 2771.56    & OOM         & OOM         & OOM       & OOM         \\ \cline{2-12}
                                  & \bf FW      & \bf 0.89 & \bf 0.44 & \bf 1.86 & \bf 2.60 & \bf 50.64 & \bf 192.31 & \bf 2044.09 & \bf 7514.67 & \bf 65.51 & \bf 4688.86 \\ \hline
    \multirow{4}{*}{\bf MetaPath} & \bf DGL     & 0.04     & 0.09     & 0.13     & 0.10     & 1.67      & 35.17      & 376.55      & OOM         & OOM       & OOM         \\ \cline{2-12}
                                  & \bf TRW     & 0.22     & 0.42     & 2.43     & 13.32    & 121.96    & 2144.53    & OOT         & OOT         & 202.27    & OOT         \\ \cline{2-12}
                                  & \bf LRW     & 0.11     & 0.19     & 0.36     & 0.61     & 9.13      & 40.24      & 422.24      & OOM         & OOM       & OOM         \\ \cline{2-12}
                                  & \bf FW      & \bf 0.01 & \bf 0.02 & \bf 0.05 & \bf 0.07 & \bf 0.65  & \bf 2.85   & \bf 37.45   & \bf 132.62  & \bf 0.98  & \bf 74.36   \\ \hline
    % \hline
    \toprule[0.8pt]
  \end{tabular}
\end{table*}

\subsection{Overall Comparison} \label{sec:overall_comparison}

Table \ref{tab:all} showcases the overall comparison of execution times across different frameworks. Notably, \textit{FW} is the only method capable of completing all test cases. In contrast, \textit{DGL}, \textit{LRW}, \textit{TRW}, and \textit{SW} struggle with larger graphs, encountering either time-out (OOT) or memory overflows (OOM). Specifically, \textit{FW} finishes all cases within merely 2.2 hours. Among scenarios where all five frameworks succeed, \textit{FW} achieves remarkable speedups. Compared with \textit{DGL} on GPU, the maximum speedup is $92.2\times$, while this number is $315.8\times$ for \textit{DGL} on CPU (executing Node2Vec). \textit{FW} reaches up to $16.4\times$, $752.2\times$ and $72.1\times$ speedup compared to \textit{LRW}, \textit{TRW} and \textit{SW} respectively, underscoring its superior performance.

\textit{FW} takes considerably longer time to process the UK, TW, and SK graphs compared to other datasets, while \textit{DGL}, \textit{LRW}, \textit{TRW}, and \textit{SW} often fail to complete within the time limit for these graphs. This increased time is attributed to the high degree of skewness in these graphs, as indicated in Table \ref{tab:data}. High-degree vertices are visited more frequently, thereby dominating the processing time. These results underscore the importance of employing different levels of samplers for vertices with varying degrees. Despite its large size, the FS graph is processed relatively quickly due to its sparsity. Although both DeepWalk and Node2Vec have the same target length, the execution time on Node2Vec is longer than that on DeepWalk because the cost of calculating the transition probability of Node2\-Vec is higher than that of DeepWalk.

\textit{FW} eliminates the need for auxiliary data structures for each query's sampling, thereby reducing the space cost per query from $O(d_{\text{max}})$ to $O(1)$, where $d_{\text{max}}$ is the maximum degree of a graph. This efficiency enables \textit{FW} to support large graphs and a substantial number of concurrent queries. \textit{FW} also exhibits superior performance on smaller graphs due to the improvement of scheduling and sampling methods. We evaluate these techniques in Section \ref{sec:detail_eval}. In summary, \textit{FW} surpasses existing CPU, GPU, and FPGA frameworks in DGRW performance and is capable of efficiently handling large graphs.

\subsection{Detailed Evaluation}\label{sec:detail_eval}

In this subsection, we have a detailed evaluation of the performance of \textit{FW}. Due to space limitations, some evaluations such as the comprehensive ablation study are provided in the technical report.

\textbf{Memory Consumption.} Table~\ref{tab:all_mem} presents a comparison of memory consumption between \textit{FW} and \textit{SW} across different datasets, with query sizes $|\mathbb{Q}|=10^6$ and $|\mathbb{Q}|=10^7$. \textit{SW} can exceed GPU memory capacity due to its use of unified virtual memory (UVM). The “extra” memory usage (\textbf{E}) is calculated by subtracting the data\-set size from the total memory consumption.

Remarkably, the extra memory consumption of \textit{FW} remains consistent across all graph sizes, whereas \textit{SW} exhibits a marked increase in memory use for larger graphs. This stability is attributable to the design of \textit{FW}. \textit{FW} minimizes per-query memory usage from $O(d)$ to $O(1)$, which is independent of graph size. It requires no auxiliary data structures in the global memory to support the execution. In contrast, \textit{SW} requires a buffer of size $O(d_{max})$ for each query and has a large task queue for load balance.

For $|\mathbb{Q}|=10^6$, query sequences occupy approximately 309 MB of memory, with a 32-bit integer representation for each vertex. For $|\mathbb{Q}|=10^7$, this figure rises to 3090 MB. Beyond storing query sequences, \textit{FW} uses no additional memory for auxiliary data structures. These findings confirm two key points: 1) existing GPU frameworks struggle with significant memory consumption issues, and 2) \textit{FW} excels in memory efficiency.

\begin{table}
  \setlength{\abovecaptionskip}{0pt}
  \setlength{\belowcaptionskip}{0pt}
  \centering
  \caption{Memory usage (GB). T refers to the total memory consumption, and E refers to the extra memory consumption (subtracting the size of dataset from T).}
  \label{tab:all_mem}
  \resizebox{0.48\textwidth}{!}{
    \begin{tabular}{c||cc|cc||cc|cc} \bottomrule[0.8pt]
      \multirow{3}{*}{\shortstack{\bf Data-                                                                                                                      \\\bf set}} & \multicolumn{4}{c||}{$|\mathbb{Q}|=10^6$} & \multicolumn{4}{c}{$|\mathbb{Q}|=10^7$} \\ \cline{2-9}
         & \multicolumn{2}{c|}{\bf FW} & \multicolumn{2}{c||}{\bf SW} & \multicolumn{2}{c|}{\bf FW} & \multicolumn{2}{c}{\bf SW}                                 \\ \cline{2-9}
         & \bf T                       & \bf E                        & \bf T                       & \bf E                      & \bf T & \bf E & \bf T & \bf E \\ \hline
      YT & 0.35                        & 0.31                         & 2.07                        & 2.02                       & 3.10  & 3.05  & 18.41 & 18.36 \\ \hline
      CP & 0.57                        & 0.30                         & 2.08                        & 1.81                       & 3.31  & 3.05  & 18.42 & 18.15 \\ \hline
      LJ & 0.96                        & 0.30                         & 2.62                        & 1.95                       & 3.71  & 3.05  & 18.96 & 18.29 \\ \hline
      OK & 2.06                        & 0.30                         & 3.81                        & 2.04                       & 4.81  & 3.05  & 20.15 & 18.38 \\ \hline
      EU & 4.24                        & 0.31                         & 8.63                        & 4.66                       & 6.99  & 3.06  & 24.97 & 21.00 \\ \hline
      AB & 8.64                        & 0.30                         & 14.32                       & 5.90                       & 11.39 & 3.05  & 30.66 & 22.24 \\ \hline
      UK & 12.12                       & 0.30                         & 26.50                       & 14.5                       & 14.87 & 3.05  & 42.83 & 30.87 \\ \hline
      TW & 18.38                       & 0.30                         & 41.60                       & 23.4                       & 21.13 & 3.05  & 57.93 & 39.70 \\ \hline
      FS & 27.46                       & 0.30                         & 29.01                       & 1.61                       & 30.21 & 3.05  & 45.35 & 17.95 \\ \hline
      SK & 27.47                       & 0.31                         & 90.99                       & 63.6                       & 30.22 & 3.06  & 107.3 & 79.98 \\
      \toprule[0.8pt]
    \end{tabular}
  }
\end{table}

\begin{figure*}[t]
  \setlength{\abovecaptionskip}{0pt}
  \setlength{\belowcaptionskip}{0pt}
  \captionsetup[subfigure]{aboveskip=0pt,belowskip=0pt}
  \centering
  \begin{subfigure}[t]{0.33\textwidth}
    \centering
    \includegraphics[width=\textwidth]{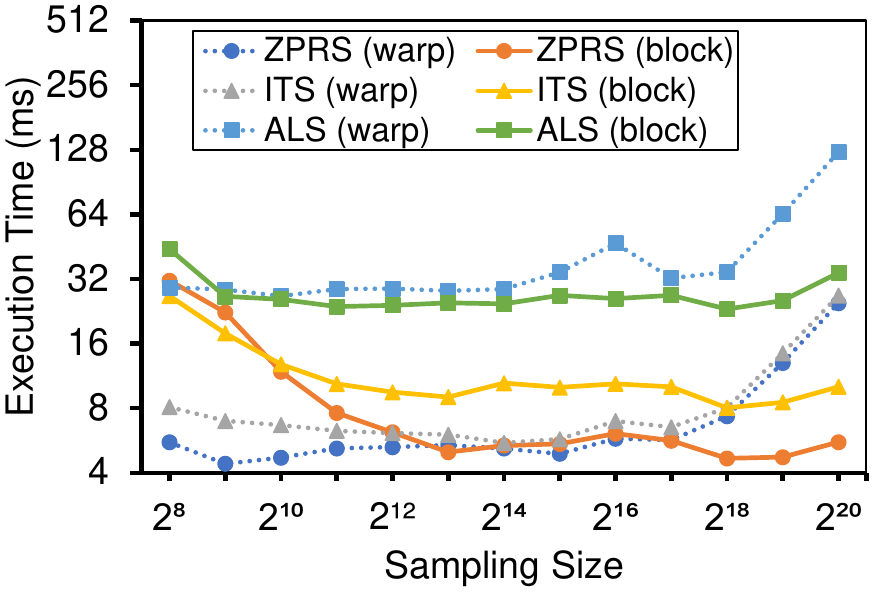}
    \caption{Comparison of ZPRS, ITS, and ALS.}
    \label{fig:sampler_time}
  \end{subfigure}
  \begin{subfigure}[t]{0.33\textwidth}
    \centering
    \includegraphics[width=\textwidth]{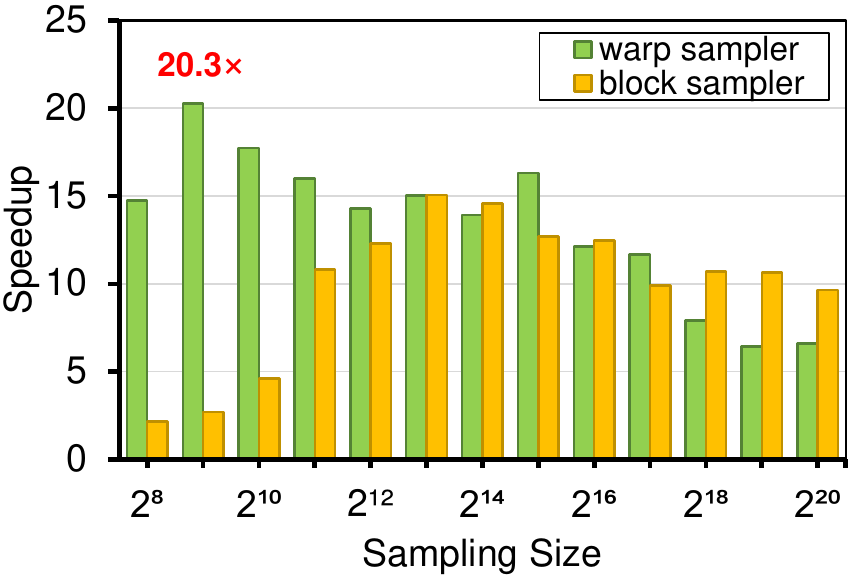}
    \caption{Impact of RNG optimization on ZPRS.}
    \label{fig:rng_speedup}
  \end{subfigure}
  \begin{subfigure}[t]{0.33\textwidth}
    \centering
    \includegraphics[width=\textwidth]{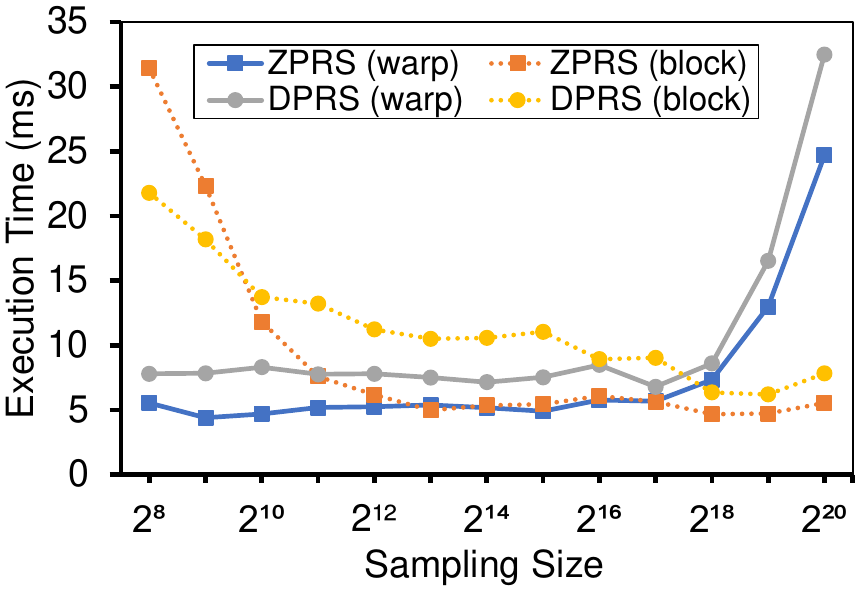}
    \caption{Comparison of ZPRS and DPRS.}
    \label{fig:two_sampler}
  \end{subfigure}
  \caption{Detailed evaluation of ZPRS and DPRS: warp and block indicate task processing at warp and block levels, respectively.}
  \label{fig:sampler}
\end{figure*}

\textbf{Evaluation of Sampling Methods.} We assess the performance of ZPRS, ITS, and ALS on GPUs by sampling a cumulative 2GB of elements, partitioned into tasks of varying sampling sizes.
“Sampling size” refers to the number of elements involved in a single sampling operation, and all tasks within a single workload share the same sampling size. Figure \ref{fig:sampler_time} reveals that ITS on warp performs comparably to ZPRS, while ZPRS on block outperforms ITS on block. This discrepancy arises because ITS necessitates frequent collective operations, which are more efficiently executed on warps than on blocks. ALS lags behind its counterparts due to the complex alias table construction. Recall that ZPRS has a space complexity of $O(1)$, while ITS has a space complexity of $O(d)$.
Our results demonstrate that ZPRS outperforms existing samplers without auxiliary data structures.

Unlike ITS, ALS, and RS, the performance of Rejection Sampling (RJS) depends on the underlying probability distribution. We observe that on less biased distributions, RJS can surpass RS at some sampling sizes due to its lower initialization cost. However, as the distribution grows more biased, RJS's performance significantly deteriorates. This variability can impact the stability of performance. Detailed results are presented in the technical report.

Figure \ref{fig:rng_speedup} highlights the substantial speedup achieved through optimizing random number generation (RNG) in ZPRS. These results underscore both the necessity and effectiveness of RNG optimization in ZPRS, as each element requires the generation of a random number. The observed speedup for ZPRS when processed on blocks is minimal for small sampling sizes because small tasks do not fully utilize the hardware capabilities. Conversely, speed gains on warps are limited for large sampling sizes, as processing extended sequences on warps does not maximize memory bandwidth utilization.

For the same reason, both ZPRS and DPRS on warps run faster than on blocks for small sampling sizes but slower for large sizes in Figure \ref{fig:two_sampler}. When the sampling size is larger than $2^9$, DPRS lags behind ZPRS for both warp and block samplers due to communication costs between threads.

\textbf{Ablation Study.} We conduct an ablation study to analyze the contributions of each individual technique to the overall speedup. Initially, we implement a baseline version of \textit{FW} with DPRS, RNGs stored in global memory, and a basic static scheduler. This setup is referred to as \textbf{FW}. Subsequently, we enhance \textbf{FW} by optimizing RNG, which we denote as \textbf{FW + RNG}. Following this, we replace DPRS with ZPRS, marked as \textbf{FW + ZPRS}. Finally, we integrate dynamic scheduling, labeled as \textbf{FW + DS}.

\begin{figure}[h]
  \setlength{\abovecaptionskip}{0pt}
  \setlength{\belowcaptionskip}{0pt}
  \captionsetup[subfigure]{aboveskip=0pt,belowskip=0pt}

  \centering
  \includegraphics[width=0.8\linewidth]{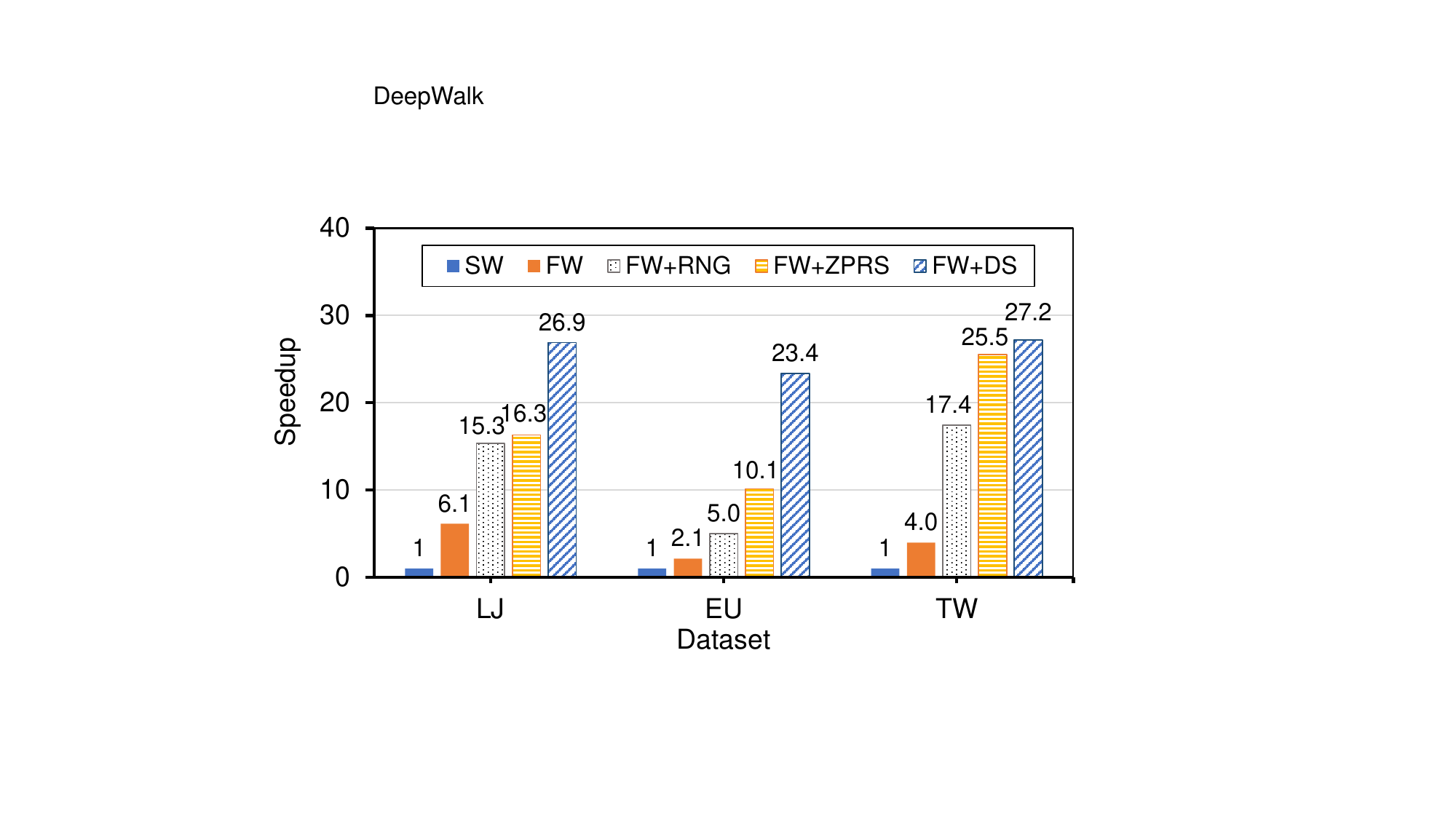}
  \caption{Speedup breakdown on DeepWalk with $10^6$ queries. The value is normalized to SkyWalker (\textit{SW}).}
  \label{fig:breakdown}
\end{figure}

The speedup of \textit{SW} on DeepWalk with $10 ^ 6$ queries against LJ, EU, and TW is illustrated in Figure \ref{fig:breakdown}. The data indicates that \textit{FW} achieves a speedup range of $2.1\times$ to $6.1\times$ over \textit{SW} without any optimizations. The optimized shared-memory RNG contributes an additional $2.5\times$ to $4.4\times$ speedup. The adoption of ZPRS further results in a speedup of $1.1\times$ to $2.0\times$. Lastly, the implementation of dynamic scheduling offers an additional $1.1\times$ to $2.3\times$ speedup. These findings affirm the efficacy of each technique introduced in our paper. It is noteworthy that ZPRS, despite being a basic operator, contributes a significant $1.1\times$ to $2.0\times$ speedup to the overall system performance. The effect of dynamic scheduling is relevant to the degree skewness of the graph. This is the reason that the speedup of \textbf{FW + DS} on TW is smaller than EU and LJ. We will elaborate on this in the technical report, as well as the ablation study results of all datasets and applications.

\subsection{Case Study}

GNNs are important for ByteDance operations, spanning video
recommendations, friend suggestions, and fraud detection. In this case study, we focus on Douyin friend recommendation scenarios.
The utilized framework is a business-specific adaptation of Graph-Learn~\cite{zhu2019aligraph}, running on a CPU-only cluster of 20 machines with 560 cores. The RW phase in the training is to perform DeepWalk where Graph-Learn executes in the dynamic mode. The test graph comprises 227 million vertices and 2.71 billion edges.

\begin{figure}[htbp]
  \setlength{\abovecaptionskip}{0pt}
  \setlength{\belowcaptionskip}{0pt}
  \captionsetup[subfigure]{aboveskip=0pt,belowskip=0pt}
  \centering
  \includegraphics[width=\linewidth]{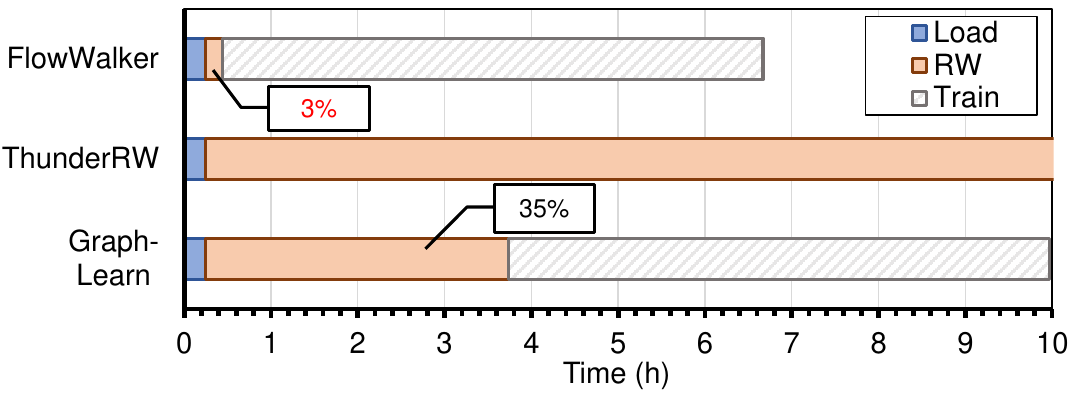}
  \caption{Time breakdown of training one epoch.}
  \label{fig:byte}
\end{figure}

Figure \ref{fig:byte} breaks down the execution time for a single training epoch. The process is composed of several key components: data loading, random walk generation, and embedding learning. Completing one epoch takes nearly 10 hours, subdivided into data loading (0.25 hours), random walk (RW) generation (3.49 hours), and network training (6.32 hours). RW occupies 35\% of the total processing time. If more advanced RW algorithms like Node2Vec are used, RW can consume much more time, as evidenced by DeepWalk vs. Node2Vec in Table \ref{tab:all}.
We do not include Node2Vec in the case study since Graph-Learn cannot support it.

As shown in Figure \ref{fig:byte}, FlowWalker reduces the RW time to merely 13 minutes (3\% of the total cycle time), offering significant efficiency gains. On the other hand, ThunderRW requires more than 10 hours on a single machine.
Skywalker is omitted from the comparison because it encounters a memory failure. These findings highlight the compelling performance advantages of Flow\-Walker.

\section{Conclusion}
In this paper, we propose FlowWalker, a memory-efficient and high-performance GPU-based framework for dynamic graph random walks. We develop DPRS and ZPRS, two parallel reservoir sampling algorithms to perform fast sampling with no extra global memory and pre-processing cost. We implement a GPU walking engine to process a massive number of walking queries based on the sampler-centric paradigm. The effectiveness of FlowWalker is evaluated through a variety of datasets, and the results show that FlowWalker achieves up to $752.2\times$ speedup on four representative random walk applications. At last, the case study reveals that FlowWalker can reduce the time cost of dynamic random walk from 35\% to 3\% of the GNN training pipeline.

% \begin{acks}
%   We sincerely thank all the anonymous reviewers for their valuable comments. This work is supported in part by the National Natural Science Foundation of China
%   (No.62122053), and a ByteDance Research Grant (CT20230525001744). The corresponding authors are Shixuan Sun and Chao Li.
%   % This work was supported by the [...] Research Fund of [...] (Number [...]). Additional funding was provided by [...] and [...]. We also thank [...] for contributing [...].
% \end{acks}

\bibliographystyle{ACM-Reference-Format}
\bibliography{main}
\balance
\clearpage

\begin{appendices}

\section*{\centering APPENDIX} % 添加跨栏的大标题

\section{Differences with Existing Memory Reduction Strategies} 
FlowWalker effectively eliminates the need for large auxiliary data structures in graph random walk processing. This approach enables the simultaneous execution of a vast number of random walks, fully leveraging GPU computing power. In contrast, GraSS \cite{grass} focuses on graph compression. However, when processing graph random walk queries, the significant overhead from per-query auxiliary data structures restricts the number of concurrent queries, leading to suboptimal utilization of computing resources. Essentially, graph compression techniques like GraSS are complementary to our work. They can be integrated with FlowWalker to further minimize memory usage. 

Training GNN on a large graph usually involves mini-batch meth\-od. Mini-batch training effectively reduces memory consumption by limiting the number of vertices in each batch. However, the sampling process for generating mini-batches in GNN, due to its stochastic nature, often requires operating on the entire graph \cite{tripathy2023distributed}. Additionally, several prominent graph learning frameworks, such as AGL \cite{agl}, adopt a two-stage processing approach. This involves initial sampling followed by mini-batch training, a method that ensures reproducibility of embedding queries and training results. The first stage entails executing a large number of random queries on the graph. It is pertinent to note that our work is primarily aimed at enhancing the efficiency of random walk queries. 

% \section{Correctness Proof}
\section{Correctness Proof of Algorithm \ref{algo:serial}} \label{sec:serial}

As the correctness proof of Algorithm \ref{algo:zigzag_parallel_rs} depends on the correctness of Algorithm \ref{algo:serial}, we first prove the correctness of Algorithm \ref{algo:serial}. The algorithm selects the element $S[i]$ with the probability $p(i) = p(i)=\frac{W[i]}{\sum^{n}_{j=1}W[j]}$ given a sequence $S$ with $n$ elements and the corresponding weight sequence $W$. We prove this using the constructive method.

\textbf{Base Case.} The algorithm apparently works for $n = 1$.

\textbf{Induction Assumption.} Given a sequence with $n$ elements and an arbitrary integer $i$ where $n > 1$ and $1 \leqslant i \leqslant n + 1$, we assume that $S[i]$ is selected with a probability of $\frac{W[i]}{\sum^{n}_{j=1}W[j]}$.

\textbf{Inductive Step.} Give a sequence with $n + 1$ elements and an arbitrary integer $i$ where $1 \leqslant i \leqslant n$, want to prove that $S[i]$ is selected with the probability of $\frac{W[i]}{\sum^{n+1}_{j=1}W[j]}$.

As shown in Line 4 in Algorithm \ref{algo:serial}, $S[n + 1]$ is chosen with a probability of $\frac{W[n+1]}{\sum^{n+1}_{j=1}W[j]}$. Thus, the current selected element has a probability of $1-\frac{W[n+1]}{\sum^{n+1}_{j=1}W[j]}=\frac{\sum^{n}_{j=1}W[j]}{\sum^{n+1}_{j=1}W[j]}$ to stay (not be replaced by element $n+1$). According to the inductive assumption, for the elements $S[i]$ positioned before $S[n + 1]$ in the sequence, $S[i]$ has the probability of $\frac{W[i]}{\sum^{n}_{j=1}W[j]}$ to be the currently selected vertex. Then, all these elements $S[i]$ has a probability of $\frac{W[i]}{\sum^{n}_{j=1}W[j]}\times\frac{\sum^{n}_{j=1}W[j]}{\sum^{n+1}_{j=1}W[j]}=\frac{W[i]}{\sum^{n+1}_{j=1}W[j]}$ to be the selected element after processing $S[n + 1]$. Thus, all the $n + 1$ elements has the probability of $\frac{W[i]}{\sum^{n+1}_{j=1}W[j]}$ to be selected. The correctness of Algorithm \ref{algo:serial} is proved. The correctness proof of ZPRS is detailed in Section \ref{sec:zprs_analysis}.

\section{Supplementary Evaluation}

\subsection{Comparison with Rejection Sampling}
In this section, we discuss and evaluate the performance of rejection sampling (RJS). Given a neighbor set $N(v)$ of a vertex $v$, RJS samples a vertex $u$ from $N(v)$ in two phases. The initialization phase calculates $p_{max} = \max_{u \in N(v)} p(u)$ where $p(u)$ is the selection probability of $u$. Subsequently, the selection phase has two steps: 1) randomly select a vertex $u$ from $N(v)$; and 2) randomly generate a real number $p$ in $[0, p_{max})$. If $p < p(u)$, then $u$ is the sampling result. Otherwise, RJS repeats the selection phase. The time complexity of initialization is $O(d(v))$, and that of selection is $O(\mathbb{E})$ where $\mathbb{E} = \frac{d(v) \times p_{max}}{\sum p(u)}$.

\begin{figure}[h]
% \setlength{\abovecaptionskip}{0pt}
%     \setlength{\belowcaptionskip}{0pt}
    % \captionsetup[subfigure]{aboveskip=0pt,belowskip=0pt}
    \centering  %居中
        % \begin{minipage}{\linewidth}
        % \centering    %子图居中
        \includegraphics[width=\linewidth]{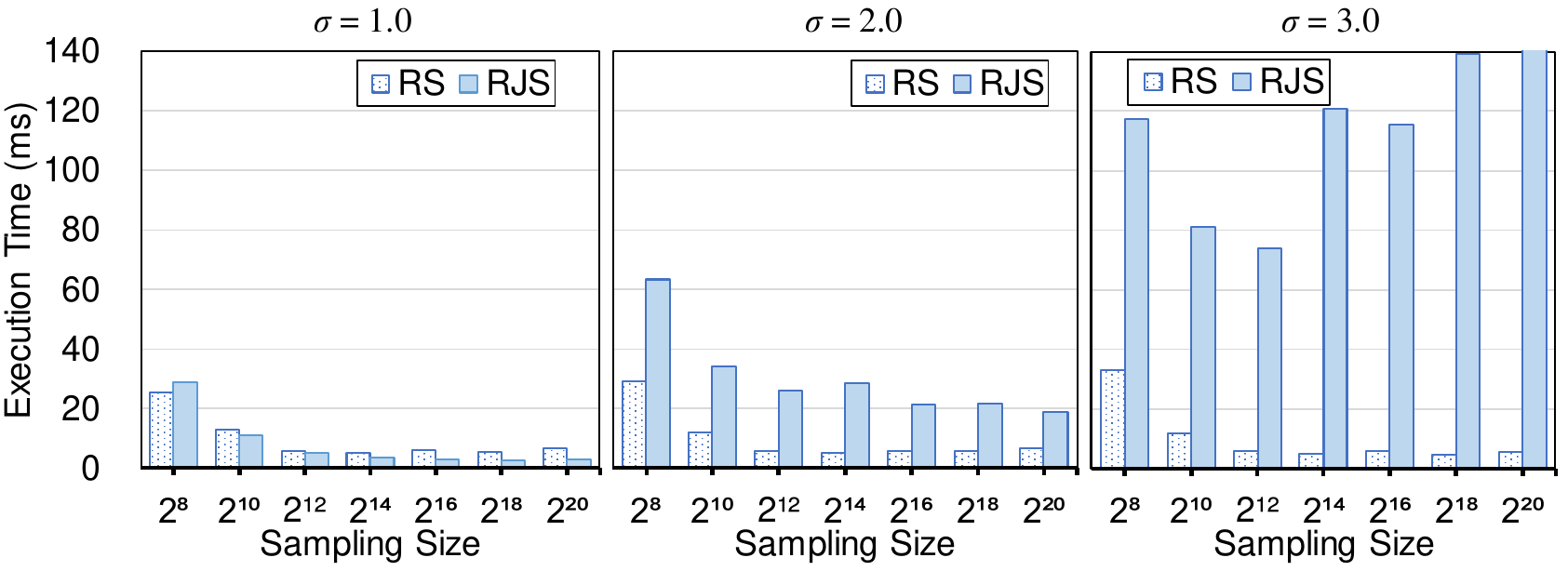}  %以pic.jpg的0.5倍大小输出
        % \end{minipage}
    \caption{Comparison of RS and RJS on different weight distributions with varying sampling sizes.}    %大图名称
    \label{fig:rjs}    %图片引用标记
\end{figure}

We empirically compare the performance of reservoir sampling (RS) with RJS under different distributions in Figure \ref{fig:rjs}. Specifically, we generate the weights using log-normal distribution with the mean value $\mu$ as 0 and the standard deviation $\sigma$ varying from 1 to 3. The sampling size is the number of elements in a single operation. When $\sigma = 1$, we can see that RS is faster than RJS on small sampling sizes (e.g., $2^8$). This is because the selection phase of RJS incurs expensive overhead. However, RS is slower on large sampling sizes (at most $2.6\times$) because 
% the initialization phase dominates the cost with the sampling size increasing and 
RS needs to generate a random number for each element that dominates the cost. When $\sigma = 2$ and $\sigma = 3$, the performance of RJS significantly degrades because the selection phase needs to be repeated many times. In contrast, the performance of RS is steady and significantly outperforms RJS (up to $39.6\times$).

\begin{table}[htbp]
\centering
\caption{Comparison of RS and RJS performance in weighted Node2Vec across various weight distributions. Values are derived using a log-norm generator with standard deviations ranging from 1 to 3.}
\label{tab:node2vec}
\resizebox{0.49\textwidth}{!}{
\begin{tabular}{c||cc||cc||cc}  \bottomrule[0.8pt]
 & \multicolumn{2}{c||}{$\sigma=1$} & \multicolumn{2}{c||}{$\sigma=2$} & \multicolumn{2}{c}{$\sigma=3$} \\ \hline
\bf Dataset & \bf RS & \bf RJS & \bf RS & \bf RJS & \bf RS & \bf RJS \\ \hline
\bf YT & \bf 0.94 & 1.02 & \bf 0.87 & 3.40 & \bf 0.76 & 8.11 \\ \hline
\bf CP & \bf 0.42 & 0.80 & \bf 0.42 & 1.44 & \bf 0.41 & 1.81 \\ \hline
\bf LJ & \bf 1.84 & 2.81 & \bf 1.72 & 8.46 & \bf 1.67 & 17.71 \\ \hline
\bf OK & \bf 2.51 & 3.62 & \bf 2.55 & 12.11 & \bf 2.50 & 30.15 \\ \hline
\bf EU & 49.87 & \bf 38.66 & \bf 49.75 & 156.25 & \bf 48.97 & 615.80 \\ \hline
\bf AB & 192.06 & \bf 133.79 & \bf 190.54 & 434.28 & \bf 186.18 & 1825.56 \\ \hline
\bf UK & 2060.20 & \bf 1251.38 & 2061.07 & \bf 1743.77 & \bf 2057.86 & 9429.69 \\ \hline
\bf TW & 7570.73 & \bf 4727.40 & 7594.72 & \bf 7303.92 & \bf 7600.89 & OOT \\ \hline
\bf FS & \bf 64.33 & 100.22 & \bf 64.09 & 360.87 & \bf 63.80 & 819.82 \\ \hline
\bf SK & 4717.55 & \bf 2834.26 & \bf 4691.10 & 4913.34 & \bf 4647.53 & 23776.17 \\ \hline
\toprule[0.8pt]
\end{tabular}
}
\end{table}

\begin{table}[t]
\centering
\caption{Comparison of RS and RJS performance in weighted MetaPath across various label distributions. Values are derived using a log-norm generator with standard deviations ranging from 1 to 3.}
\label{tab:metapath}
\resizebox{0.48\textwidth}{!}{
\begin{tabular}{c||cc||cc||cc}  \bottomrule[0.8pt]
 & \multicolumn{2}{c||}{$\sigma=1$} & \multicolumn{2}{c||}{$\sigma=2$} & \multicolumn{2}{c}{$\sigma=3$} \\ \hline
\bf Dataset & \bf RS & \bf RJS & \bf RS & \bf RJS & \bf RS & \bf RJS \\ \hline
\bf YT & \bf 0.01 & 0.01 & \bf 0.01 & 0.04 & \bf 0.01 & 0.09 \\ \hline
\bf CP & \bf 0.02 & 0.02 & \bf 0.02 & 0.03 & \bf 0.02 & 0.03 \\ \hline
\bf LJ & \bf 0.04 & 0.05 & \bf 0.04 & 0.12 & \bf 0.04 & 0.22 \\ \hline
\bf OK & \bf 0.06 & 0.09 & \bf 0.06 & 0.25 & \bf 0.06 & 0.52 \\ \hline
\bf EU & \bf 0.63 & 0.69 & \bf 0.62 & 3.52 & \bf 0.62 & 12.06 \\ \hline
\bf AB & \bf 1.54 & 2.63 & \bf 1.51 & 13.03 & \bf 1.50 & 51.59 \\ \hline
\bf UK & 77.80 & \bf 24.22 & 78.59 & \bf 54.19 & \bf 78.08 & 348.52 \\ \hline
\bf TW & 87.72 & \bf 85.72 & \bf 86.71 & 227.87 & \bf 86.53 & 1448.48 \\ \hline
\bf FS & \bf 1.32 & 1.73 & \bf 1.32 & 4.42 & \bf 1.31 & 7.28 \\ \hline
\bf SK & \bf 40.93 & 48.41 & \bf 40.35 & 172.70 & \bf 39.92 & 912.44 \\ \hline
\toprule[0.8pt]
\end{tabular}
}
\end{table}

In summary, the running time of RJS is non-deterministic and heavily depends on the probability distribution. This can affect the system stability given the complex real-world scenarios, e.g., Table \ref{tab:node2vec} and Table \ref{tab:metapath} present experiment results on weighted Node2Vec and weighted MetaPath with different distributions. Nevertheless, an interesting research direction is to dynamically select the sampling method given the input. However, this requires an efficient and effective adaptive sampling method selection mechanism at runtime, which we will leave as a future work.

\subsection{Impact of Hyperparameters} \label{sec:param}
\begin{figure}[t]
% \setlength{\abovecaptionskip}{0pt}
%     \setlength{\belowcaptionskip}{0pt}
    % \captionsetup[subfigure]{aboveskip=0pt,belowskip=0pt}
    \centering
    \begin{subfigure}[t]{0.23\textwidth}
        \centering    %子图居中
        \includegraphics[width=\linewidth]{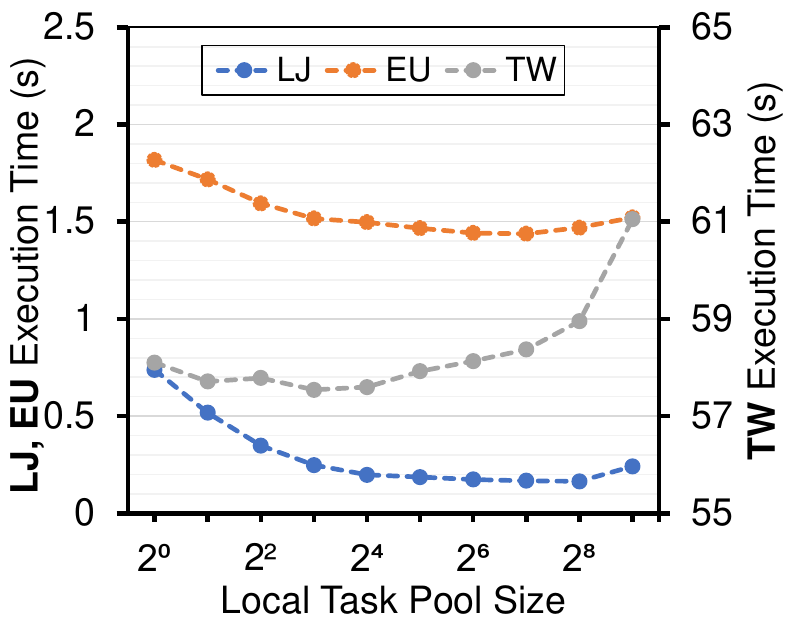}  %以pic.jpg的0.5倍大小输出
        % \end{minipage}
        \caption{Local task pool size.}    %大图名称
        \label{fig:para-size}    %图片引用标记
    \end{subfigure}
    \begin{subfigure}[t]{0.23\textwidth}
        \centering    %子图居中
         \includegraphics[width=\linewidth]{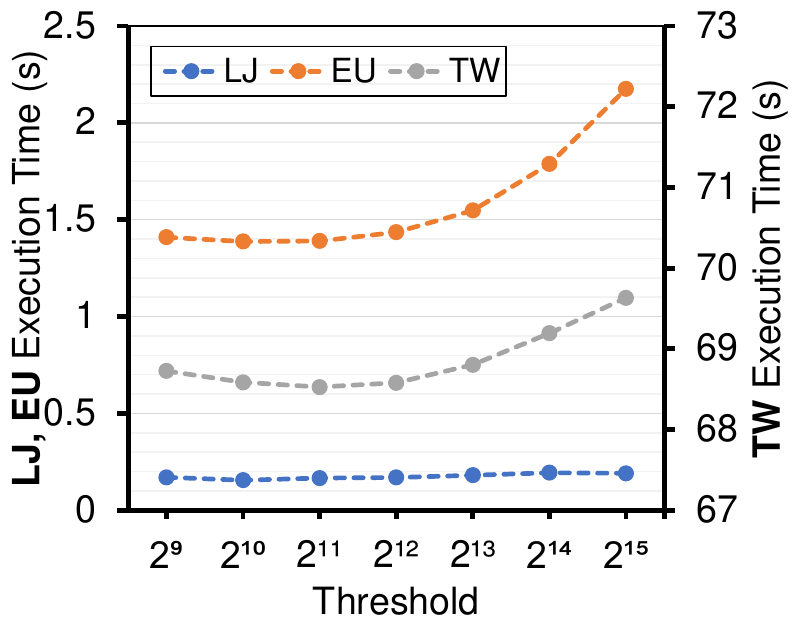}  %以pic.jpg的0.5倍大小输出
        % \end{minipage}
        \caption{Warp-block threshold.}    %大图名称
        \label{fig:para-threshold}    %图片引用标记
    \end{subfigure}
    \caption{Impact of hyperparameters on the performance.}    %大图名称
    \label{fig:hyperparameter}    %图片引用标记
\end{figure}

FlowWalker requires two hyperparameters: the local task pool size $|P_L|$, and the degree threshold $d_t$. $|P_L|$ dictates the number of queries that a thread block can hold. A small $|P_L|$ would underutilize the computational resources, while a large $|P_L|$ would lead to coarse-grained task fetching and may aggravate load imbalance. Besides, as GPUs have limited shared memory sizes, a large $|P_L|$ can exceed the shared memory limitations. Following this guideline, we can tune $|P_L|$ by varying it from $1$ to $|T|$ where $T$ is the number of threads in a block. 
% We select $\frac{|T|}{32}$ and $|T|$ as the minimum and maximum values of the range, respectively. This choice is based on two considerations: 1) a warp, consisting of 32 threads, is the basic execution unit in GPU architectures; and 2) the shared memory in GPUs is limited in size. Figure \ref{fig:para-size} displays the experimental results with $|P_L|$ varying from 32 to 1024.
Figure \ref{fig:para-size} displays the experimental results with $|P_L|$ varying from 1 to 512. It is observed that the execution time remains relatively stable within the range of $\left[ 2^5, 2^8 \right]$, but exhibits a slight increase beyond this interval. Consequently, the performance of FlowWalker demonstrates robustness to changes in $|P_L|$, leading us to set 64 as the default value for all datasets.

The parameter $d_t$ serves as the threshold for selecting between the warp sampler and block sampler. To assess the impact of this threshold on performance, we conducted micro-benchmarking experiments. The results are depicted in Figure \ref{fig:para-threshold}. Our observations reveal that the execution time remains consistent when $d_t$ ranges from $2^9$ to $2^{12}$, showing a slight increase for values of $d_t$ larger than this range. Consequently, FlowWalker exhibits robust performance relative to variations in $d_t$. Based on these findings, we have chosen $1024$ as the default value for $d_t$.

In summary, the performance of FlowWalker is robust with respect to the hyperparameters. Users can adhere to the default settings for optimal performance. In light of the reviewer's comments, we expand our discussion in Section 6.1 in the revision to include guidelines on parameter settings and their impacts.

\subsection{GPU Resource Utilization}
\begin{figure}[htbp]
% \setlength{\abovecaptionskip}{0pt}
%     \setlength{\belowcaptionskip}{0pt}
    % \captionsetup[subfigure]{aboveskip=0pt,belowskip=0pt}
    \centering  %居中
    % \subfigure[Maximum SM utilization.]{   %第一张子图
    %     \begin{minipage}{0.47\linewidth}
    \begin{subfigure}[t]{0.23\textwidth}
        \centering    %子图居中
        \includegraphics[width=\linewidth]{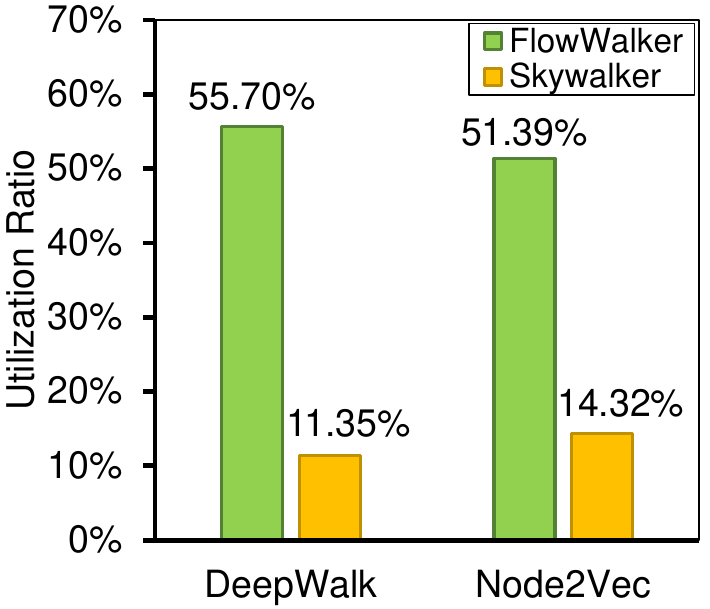}  %以pic.jpg的0.5倍大小输出
        \caption{Maximum SM utilization.}
        \label{fig:profile_sm}
    \end{subfigure}
    %     \end{minipage}
    % }
    % \subfigure[Memory bandwidth.]{   %第一张子图
    %     \begin{minipage}{0.47\linewidth}
    \begin{subfigure}[t]{0.23\textwidth}
        \centering    %子图居中
        \includegraphics[width=\linewidth]{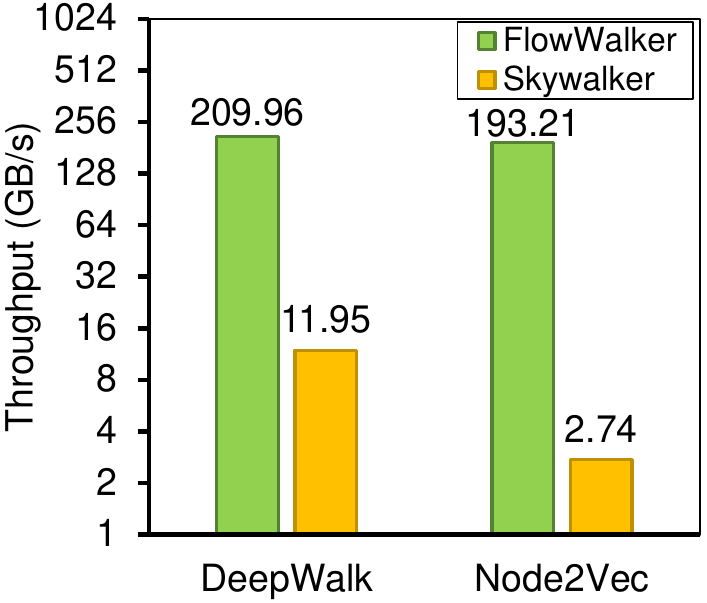}  %以pic.jpg的0.5倍大小输出
        \caption{Memory bandwidth.}
        \label{fig:profile_mb}
    \end{subfigure}
    %     \end{minipage}
    % }
    \caption{Comparison of GPU resource utilization on LJ.}    %大图名称
    \label{fig:profile}    %图片引用标记
\end{figure}

% \textbf{GPU Resource Utilization.} 
We analyze GPU resource utilization using \emph{NVIDIA Nsight Compute}. Figure \ref{fig:profile} showcases the performance metrics of DeepWalk and Node2Vec on the LJ dataset. As evidenced in Figure \ref{fig:profile_sm}, \textit{FW} boasts substantially higher SM (Streaming Multiprocessor) utilization compared to \textit{SW}, highlighting superior parallelism of \textit{FW}. Additionally, Figure \ref{fig:profile_mb} reveals that \textit{FW} enjoys a significantly higher memory bandwidth, which is attributed to its efficient coalesced memory access.

For \textit{FW}, both the SM utilization and memory bandwidth are marginally lower for Node2Vec than for DeepWalk. This is due to the binary search operations required for transition probability calculation, which lead to some random memory accesses. Despite this, \textit{FW} still outperforms \textit{SW} in overall resource utilization, demonstrating the efficacy of our approach.

\subsection{RNG Performance Evaluation}
Figure \ref{fig:rng_speedup} highlights the substantial speedup achieved through optimizing random number generation (RNG) in ZPRS. These results underscore both the necessity and effectiveness of RNG optimization in ZPRS, as each element requires the generation of a random number. The observed speedup for ZPRS when processed on blocks is minimal for small sampling sizes because small tasks do not fully utilize the hardware capabilities. Conversely, speed gains on warps are limited for large sampling sizes, as processing extended sequences on warps does not maximize memory bandwidth utilization.

\begin{figure}[htbp]
\setlength{\abovecaptionskip}{0pt}
    \setlength{\belowcaptionskip}{0pt}
    \captionsetup[subfigure]{aboveskip=0pt,belowskip=0pt}
    \centering  %居中
        % \begin{minipage}{\linewidth}
        % \centering    %子图居中
        \includegraphics[width=0.9\linewidth]{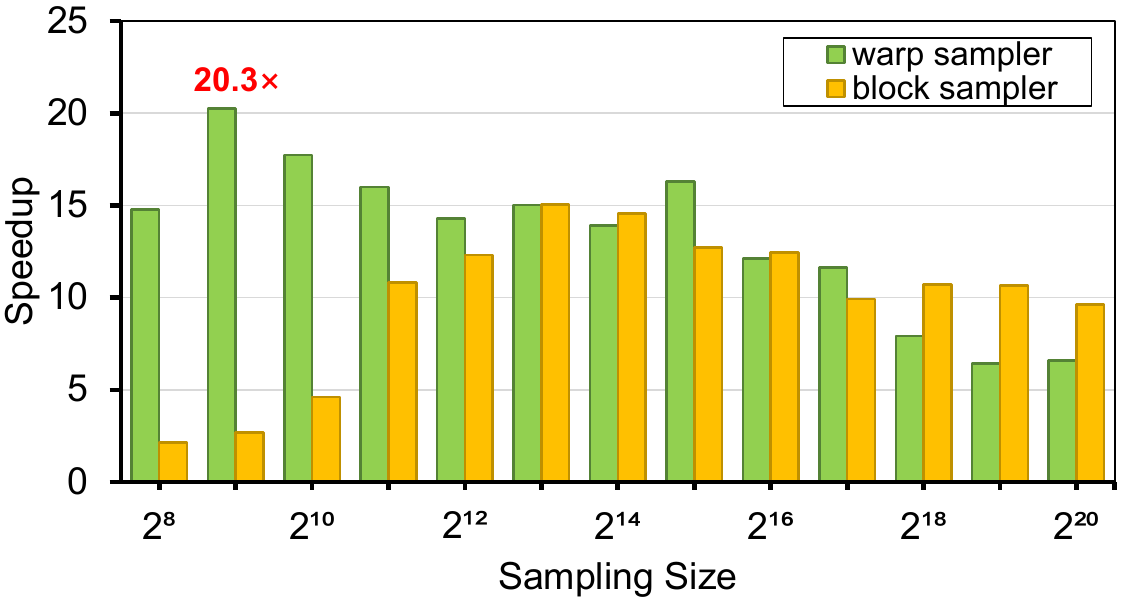}  %以pic.jpg的0.5倍大小输出
        % \end{minipage}
    \caption{Impact of RNG optimization on ZPRS.}    %大图名称
    \label{fig:rng_speedup}    %图片引用标记
\end{figure}

\balance
\subsection{Scalability Evaluation}
% \sun{1. Varying the length of random walks. 2. Varying the number of queries. Cover the cases that muliple-batches are used. 3. Varying the dataset to show that the memory consumption is independent of the graph size (except the graph storage).}

We assess the scalability of FlowWalker by examining throughput under varying numbers of walking queries and walking lengths. Specifically, throughput is defined as the number of edges processed per second. We opt for this metric over the number of vertices processed to avoid bias introduced by degree skewness. By default, we conduct $10^6$ walking queries starting from randomly selected vertices with a walking length of 80.

\begin{figure}[htbp]
    \captionsetup[subfigure]{aboveskip=0pt,belowskip=0pt}
    \centering  %居中
    \begin{subfigure}[t]{0.23\textwidth}
    % \subfigure[Varying number of queries.]{   %第一张子图
    %     \begin{minipage}{0.47\linewidth}
        \centering    %子图居中
        \includegraphics[width=\linewidth]{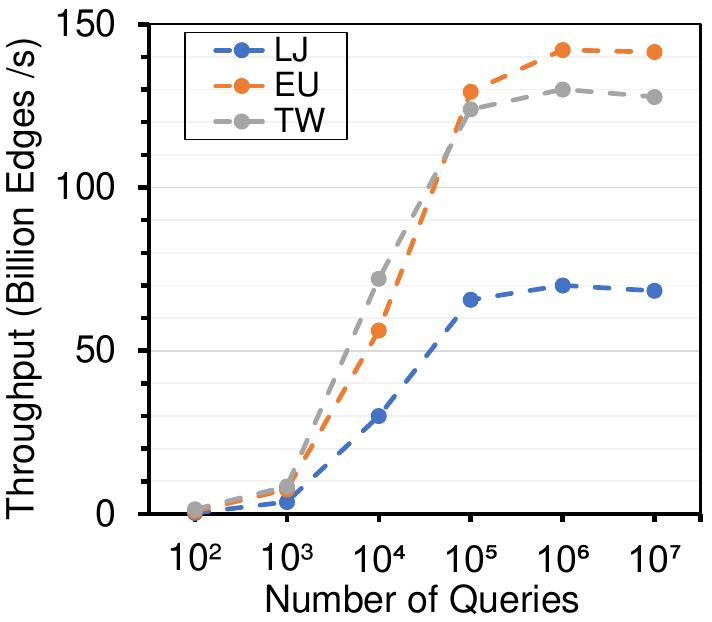}  %以pic.jpg的0.5倍大小输出
        \caption{Varying number of queries.}
        \label{fig:scale_n}
    %     \end{minipage}
    % }
    \end{subfigure}
    \begin{subfigure}[t]{0.23\textwidth}
    % \subfigure[Varying length of queries.]{   %第一张子图
    %     \begin{minipage}{0.47\linewidth}
        \centering    %子图居中
        \includegraphics[width=\linewidth]{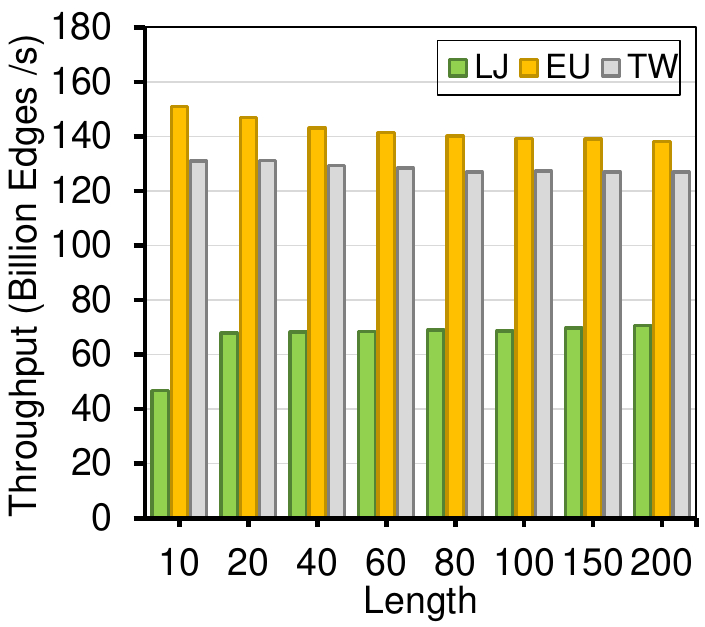}  %以pic.jpg的0.5倍大小输出
        \caption{Varying length of queries.}
        \label{fig:scale_d}
    %     \end{minipage}
    % }
    \end{subfigure}
    \caption{Throughput of \textit{FW} with varying walker number and query length.}    %大图名称
    \label{fig:scale}    %图片引用标记
\end{figure}

% \textbf{Varying walking queries number and length.} 
Figure \ref{fig:scale_n} illustrates how throughput varies as the number of queries changes from $10^2$ to $10^7$. The throughput is suboptimal at low query counts because the workload is insufficient to fully utilize the computational capacity of the GPUs. It plateaus at around $10^6$ queries, indicating strong scalability in relation to the number of queries.

In Figure \ref{fig:scale_d}, the throughput stabilizes when the query length exceeds 20, confirming the system scalability with respect to query length. During these experiments, we observed significantly higher throughput on the TW and EU datasets compared to LJ. This can be attributed to the degree distribution of tasks: over 97\% of tasks on LJ have small degrees. Consequently, the system efficiency is lower when processing a large volume of tasks that each involves scanning only a short sequence of neighbors, as opposed to EU and TW having many tasks that each scan a longer sequence.

% For LJ, the throughput increases with the length varying from 10 to 20 and then maintains at a level. And the throughput remains at the same level for different query lengths on the other two datasets. The results demonstrate the generality of \textit{FW} and that the throughput is relevant to the feature of datasets.
%The throughput of TW remains stable with varying query lengths. And for LJ, the throughput increases with the length varies from 10 to 20 and then maintains at a level.

% \begin{figure}[htbp]
%     \centering  %居中
%         % \begin{minipage}{\linewidth}
%         % \centering    %子图居中
%         \includegraphics[width=\linewidth]{figures/scale_mem.pdf}  %以pic.jpg的0.5倍大小输出
%         % \end{minipage}
%     \caption{Extra memory cost with varying datasets.}    %大图名称
%     \label{fig:scale_mem}    %图片引用标记
% \end{figure}

% \textbf{Memory cost of varying datasets.} Fig.\ref{fig:scale_mem} shows the additional memory cost variation of all testing datasets. The extra memory cost of \textit{FW} mainly comes from recording the walking path, and the sampling process does not produce any intermediate data that has to be stored in global memory. The extra memory space remains constant on different datasets. It is independent of dataset, but relevant to the number of walking queries. While for \textit{SW}, the additional memory cost also depends on the buffer allocated according to the graph.

\begin{figure*}[t]
% \setlength{\abovecaptionskip}{0pt}
%     \setlength{\belowcaptionskip}{0pt}
    % \captionsetup[subfigure]{aboveskip=0pt,belowskip=0pt}
    % \renewcommand{\thefigure}{\Roman{figure}}
    \centering
    \begin{subfigure}[t]{0.47\textwidth}
        \centering    %子图居中
        \includegraphics[width=\linewidth]{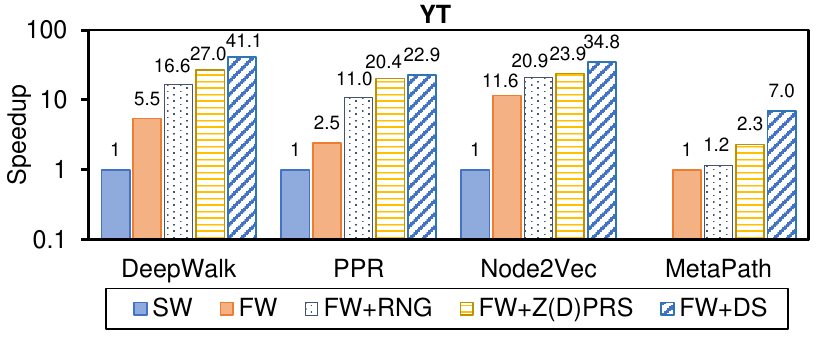}  %以pic.jpg的0.5倍大小输出
        % \end{minipage}
        \caption{Ablation study on YT.}    %大图名称
        \label{fig:yt}    %图片引用标记
    \end{subfigure}
    \begin{subfigure}[t]{0.47\textwidth}
        \centering    %子图居中
          \includegraphics[width=\linewidth]{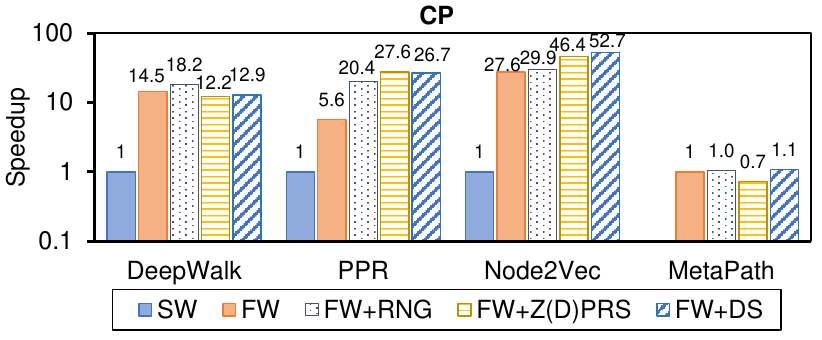}  %以pic.jpg的0.5倍大小输出
        % \end{minipage}
        \caption{Ablation study on CP.}    %大图名称
        \label{fig:cp}    %图片引用标记
    \end{subfigure}
    \begin{subfigure}[t]{0.47\textwidth}
        \centering    %子图居中
          \includegraphics[width=\linewidth]{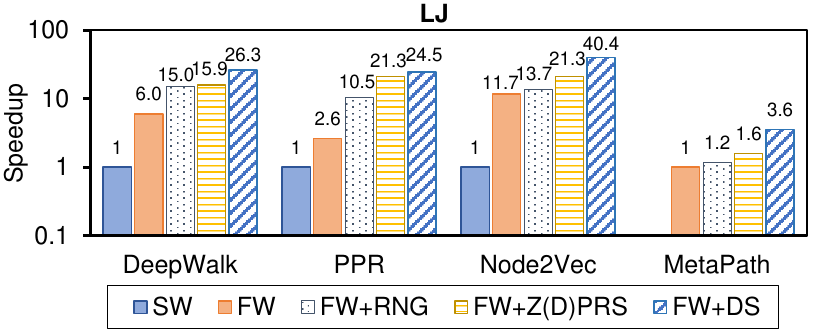}  %以pic.jpg的0.5倍大小输出
        % \end{minipage}
        \caption{Ablation study on LJ.}    %大图名称
        \label{fig:lj}    %图片引用标记
    \end{subfigure}
    \begin{subfigure}[t]{0.47\textwidth}
        \centering    %子图居中
          \includegraphics[width=\linewidth]{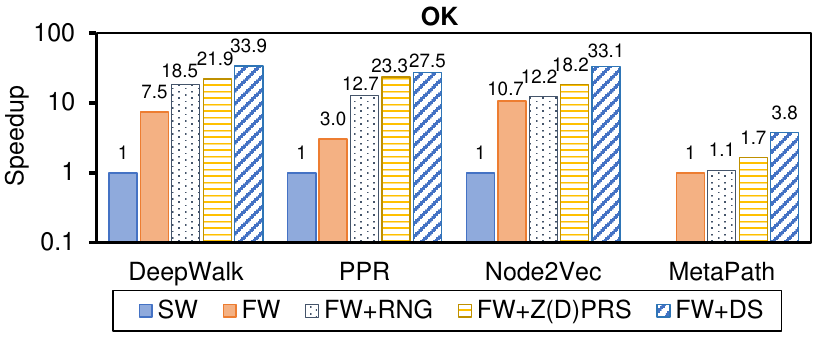}  %以pic.jpg的0.5倍大小输出
        % \end{minipage}
        \caption{Ablation study on OK.}    %大图名称
        \label{fig:ok}    %图片引用标记
    \end{subfigure}
    \begin{subfigure}[t]{0.47\textwidth}
        \centering    %子图居中
          \includegraphics[width=\linewidth]{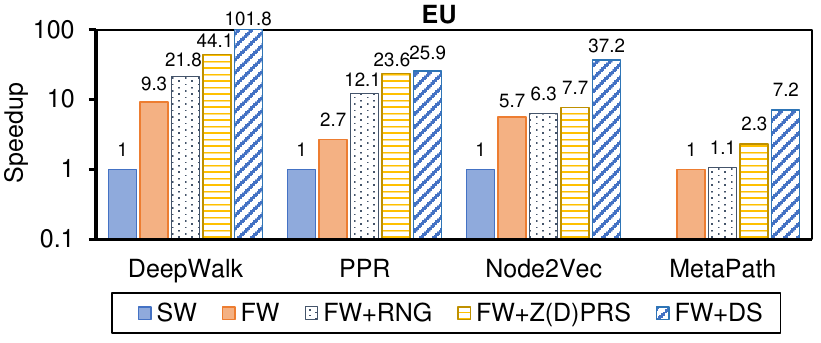}  %以pic.jpg的0.5倍大小输出
        % \end{minipage}
    \caption{Ablation study on EU.}    %大图名称
    \label{fig:eu}    %图片引用标记
    \end{subfigure}
    \caption{The results of ablation study on small and medium graphs. For DeepWalk, PPR, and Node2Vec, the results are normalized to Skywalker (SW). For MetaPath, we normalize the results to FW as SW does not support MetaPath.}    %大图名称
    \label{fig:small}    %图片引用标记
\end{figure*}

\begin{figure*}[t]
% \setlength{\abovecaptionskip}{0pt}
%     \setlength{\belowcaptionskip}{0pt}
    % \captionsetup[subfigure]{aboveskip=0pt,belowskip=0pt}
    % \renewcommand{\thefigure}{\Roman{figure}}
    \centering
    \begin{subfigure}[t]{0.47\textwidth}
        \centering    %子图居中
        \includegraphics[width=\linewidth]{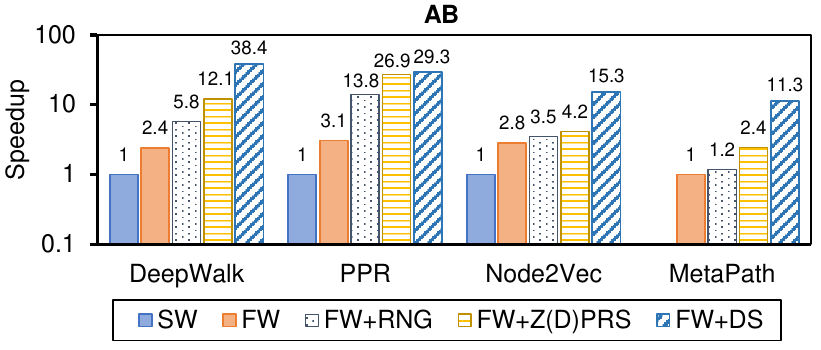}  %以pic.jpg的0.5倍大小输出
        % \end{minipage}
        \caption{Ablation study on AB.}    %大图名称
        \label{fig:ab}    %图片引用标记
    \end{subfigure}
    \begin{subfigure}[t]{0.47\textwidth}
        \centering    %子图居中
          \includegraphics[width=\linewidth]{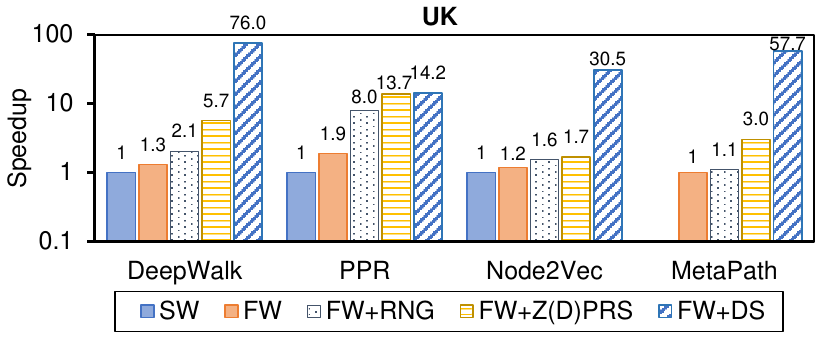}  %以pic.jpg的0.5倍大小输出
        % \end{minipage}
        \caption{Ablation study on UK.}    %大图名称
        \label{fig:uk}    %图片引用标记
    \end{subfigure}
    \begin{subfigure}[t]{0.47\textwidth}
        \centering    %子图居中
          \includegraphics[width=\linewidth]{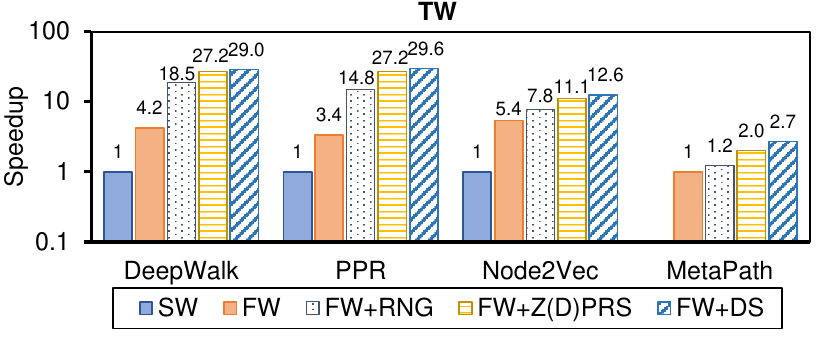}  %以pic.jpg的0.5倍大小输出
        % \end{minipage}
        \caption{Ablation study on TW.}    %大图名称
        \label{fig:tw}    %图片引用标记
    \end{subfigure}
    \begin{subfigure}[t]{0.47\textwidth}
        \centering    %子图居中
          \includegraphics[width=\linewidth]{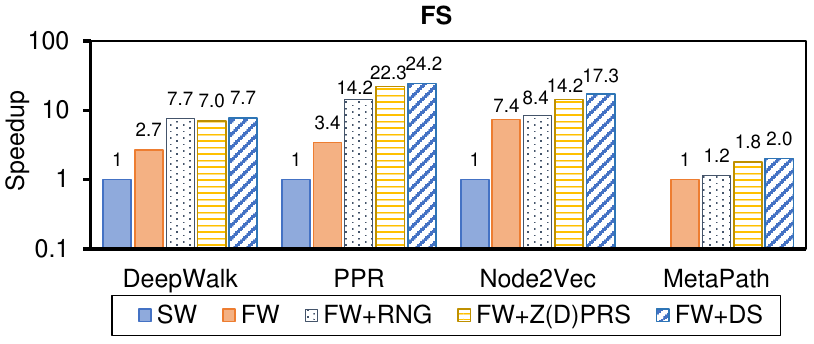}  %以pic.jpg的0.5倍大小输出
        % \end{minipage}
        \caption{Ablation study on FS.}    %大图名称
        \label{fig:fs}    %图片引用标记
    \end{subfigure}
    \begin{subfigure}[t]{0.47\textwidth}
        \centering    %子图居中
          \includegraphics[width=\linewidth]{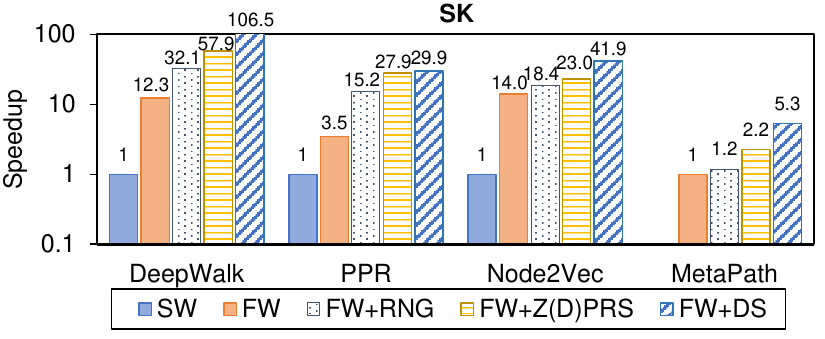}  %以pic.jpg的0.5倍大小输出
        % \end{minipage}
        \caption{Ablation study on SK.}    %大图名称
        \label{fig:sk}    %图片引用标记
    \end{subfigure}
    \caption{The results of ablation study on five billion-scale graphs.}    %大图名称
    \label{fig:large}    %图片引用标记
\end{figure*}

% \balance
\subsection{Ablation Study}
We conduct the ablation study on all datasets and four applications, assessing the impact of each technique individually. Notice that the original Skywalker code has bugs leading to fewer sampling steps. We add some thread fence to eliminate this problem\footnote{\url{https://github.com/junyimei/Skywalker}} in this section. However, the Skywalker we used as the baseline in the paper is in its original state without any modification\footnote{\url{https://github.com/wpybtw/Skywalker}}. The findings reveal that ZPRS contributes to speedups of $1.1\times$-$2.8\times$, while the implementation of DS leads to speedups of $1.03\times$-$19.0\times$. Notably, DPRS is utilized by default in Node2Vec to mitigate the transition probability computation overhead, resulting in DPRS outperforming ZPRS by $1.1\times$ to $1.7\times$. This outcome validates our cost analysis for DPRS and ZPRS, underscoring the significance of both techniques. Specifically, on five billion-scale graphs—AB, UK, TW, FS, and SK—ZPRS and DS facilitate up to $2.8\times$ and $19.0\times$ speedup, respectively, showcasing their efficiency on large-scale graph datasets. The results demonstrate the effectiveness of the techniques proposed in this paper.

We develop a baseline version of FlowWalker (\textbf{FW}), featuring DPRS, RNGs in global memory, and a simple static scheduler. This configuration is identified as \textbf{FW}. We then improved \textbf{FW} by optimizing RNG storage, creating the variant \textbf{FW + RNG}. Next, we substituted DPRS with ZPRS. For Node2Vec, we incorporate ZPRS in \textbf{FW}, and replace it with DPRS in this step. Therefore we denote this configuration as \textbf{FW + Z(D)PRS}. Finally, we incorporated dynamic scheduling, producing \textbf{FW + DS}. For Node2Vec experiments, \textbf{FW} initially uses ZPRS.

Figure \ref{fig:small} and Figure \ref{fig:large} depict the results of all the datasets listed in Table \ref{tab:data}. The results are normalized to Skywalker (SW). Specifically, \textit{for MetaPath, we normalize the results to \textbf{FW} as SW does not support MetaPath}. According to the data, the proposed optimization methods in FlowWalker are able to enhance performance in the majority of scenarios (146 of 150 cases), and FlowWalker (\textbf{FW+DS}) outperforms SW on all cases. The performance varies across different datasets and different applications. We broadly summarize several general patterns as follows.

The performance of \textbf{FW} surpasses SW on all datasets without any optimizations, including the shared-memory RNG. And the speedup is substantial on many datasets. Especially for the application Node2Vec, the baseline \textbf{FW} can provide up to $27.6\times$ speedup, which is much higher than the speedup from RNG (up to $1.8\times$), DPRS (up to $1.7\times$), and dynamic scheduling (up to $18.2\times$). This demonstrates the effectiveness of reservoir sampling.

The optimized RNG improves performance in all scenarios, but it is not the main contribution to the enhancement. The performance gain with RNG of PPR is higher than the other three applications, with a range of $3.6\times$ to $4.8\times$ speedup. This is because the PPR queries start from the vertex with the highest degree in the experiment setting. The average amount of random numbers generated is higher than the other applications. In terms of MetaPath, RNG contributes up to $1.3\times$ speedup to the overall performance. This is because in MetaPath, edge labels have to match the given pattern. FlowWalker only needs to generate random numbers for the matched edges.

\begin{figure*}[t]
% \setlength{\abovecaptionskip}{0pt}
%     \setlength{\belowcaptionskip}{0pt}
    % \captionsetup[subfigure]{aboveskip=0pt,belowskip=0pt}
    % \renewcommand{\thefigure}{\Roman{figure}}
    \centering
    \begin{subfigure}[t]{0.35\textwidth}
        \centering    %子图居中
        \includegraphics[width=\linewidth]{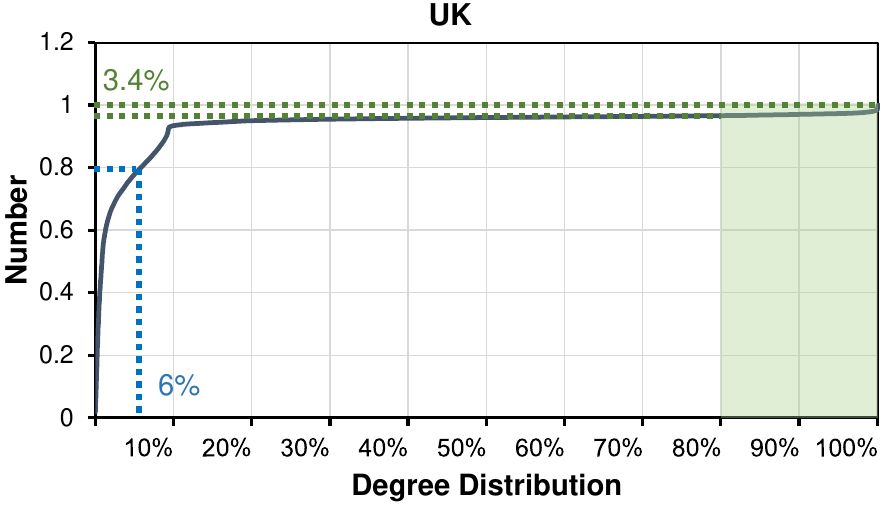}  %以pic.jpg的0.5倍大小输出
        % \end{minipage}
        \caption{Degree CDF of UK.}    %大图名称
        \label{fig:cdf_uk}    %图片引用标记
    \end{subfigure}
    \hspace{17mm}
    \begin{subfigure}[t]{0.35\textwidth}
        \centering    %子图居中
        \includegraphics[width=\linewidth]{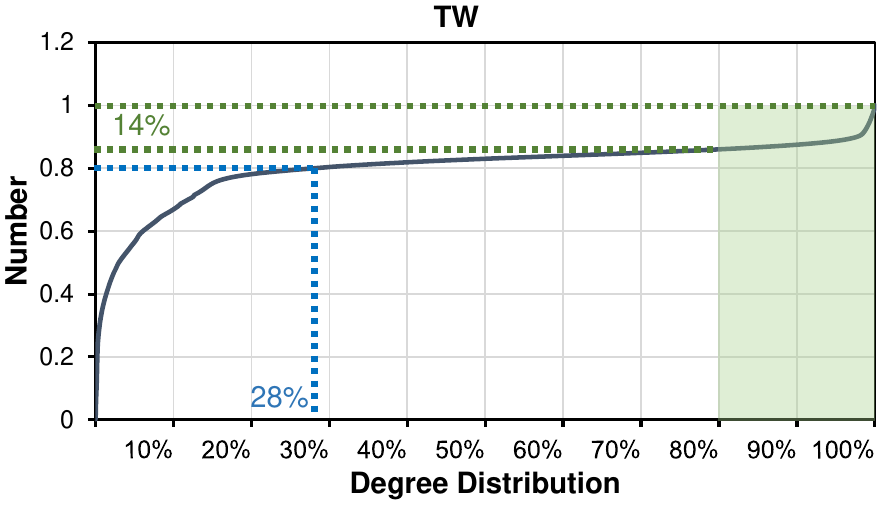}  %以pic.jpg的0.5倍大小输出
        % \end{minipage}
        \caption{Degree CDF of TW.}   %大图名称
        \label{fig:cdf_tw}    %图片引用标记
    \end{subfigure}
    \caption{The degree distribution differs in UK and TW.}    %大图名称
    \label{fig:cdf}    %图片引用标记
\end{figure*}
ZPRS contributes $1.1\times$ to $2.8\times$ speedup to the overall system performance. For Node2Vec, DPRS provides $1.1\times$ to $1.7\times$ speedup. The improvement is non-trivial for a primitive operator.

Dynamic scheduling enhances the system in most cases (149 of 150 cases), especially on skewed datasets. EU, AB, and UK greatly benefit from the dynamic scheduling method. For the highly skewed dataset UK, the speedup reaches at most $19.0\times$. The performance rise of dynamic scheduling is relevant to the workload. For example, in our experiment setting of PPR, all queries start from the same vertex with the highest degree, and there is a minor discrepancy between the workloads. Therefore the speedup of PPR is not as high as other applications, and for dataset CP, the performance even drops by 3\%. Despite this, dynamic scheduling is still an effective optimization approach.

Next, we showcase the ablation study results on five billion-scale graphs and analyze the dynamic scheduling performance on TW. The graphs include AB (Figure \ref{fig:ab}), UK (Figure \ref{fig:uk}), TW (Figure \ref{fig:tw}), FS (Figure \ref{fig:fs}), and SK (Figure \ref{fig:sk}). ZPRS (DPRS in Node2Vec) facilitates up to $2.1\times$, $2.8\times$, $1.7\times$, $1.7\times$, and $1.9\times$ improvement for five datasets respectively. The results demonstrate the effectiveness of ZPRS on the billion-scale graphs. Dynamic scheduling offers up to $4.7\times$, $19.0\times$, $1.4\times$, $1.2\times$, and $2.4\times$ speedup on these datasets.

As discussed above, the effectiveness of dynamic scheduling varies among different datasets, depending on the graph degree distribution. Take the datasets UK and TW as two examples. Figure \ref{fig:cdf_uk} and Figure \ref{fig:cdf_tw} depict the CDF curves of their degree distributions. The x-axis in the graph represents the degree distribution, for example, $10\%$ stands for the vertices with the least $10\%$ degrees. And the y-axis depicts the vertex numbers. UK is a highly skewed graph with 80\% of the vertices having the lowest 10\% degrees, and the vertices with the highest 20\% degrees take up only 3.4\% of all the nodes. Compared with UK, the degree distribution of TW is more uniform, with 80\% vertices having degrees less than 28\%, and 14\% vertices fall into the degree range of 80\% - 100\%. The skewed workload brought by skewed degree distribution in UK incurs significant performance rise of dynamic scheduling. In contrast, the speedup for TW is marginal due to the uniform workload.

The optimization techniques can lead to negative speedup in a minority of scenarios (4 of 150 cases). ZPRS incurs a performance drop when computing DeepWalk on CP (about 30\%) and FS (about 10\%). This is because CP and FS are sparse graphs, with a maximum degree of 793 and 5214 respectively. This number is much smaller than the other graphs. As shown in Figure \ref{fig:two_sampler}, the performance of DPRS is better than ZPRS when the vertex degree is small. The case is the same for MetaPath on CP. Besides, as described above, dynamic scheduling slightly degrades the performance with 3\% when performing MetaPath on CP. Since dynamic scheduling incurs additional overhead compared with the static method. The performance gain in other cases can offset the overhead, but for dataset CP, which is a small and sparse graph, dynamic scheduling degrades the performance.

In summary, the optimization methods in FlowWalker are able to enhance performance across the majority of scenarios. Combining all optimizations, FlowWalker significantly outperforms its counterpart in all cases. It is noteworthy that ZPRS contributes $1.1\times$ to $2.8\times$ speedup to the overall system performance. In particular, ZPRS contributes $1.7\times$-$2.8\times$ speedup on billion-scale graphs. The improvement is non-trivial for a primitive operator. Moreover, the speedup varies across different datasets and applications. Each acceleration technique possesses a unique zone of superiority, manifesting divergent outcomes across various contexts. Consequently, within FlowWalker, the incorporation of each optimization approach is deemed essential. 
\balance
\end{appendices}

\end{document}